\documentclass[aps,footinbib,showpacs,twocolumn,showkeys,amssymb,amsmath,superscriptaddress]{revtex4}

\usepackage{braket}
\usepackage{color}
\usepackage{stackengine}
\usepackage{verbatim}
\usepackage{graphicx}
\usepackage{xcolor}

\usepackage{bm}
\usepackage{latexsym}
\usepackage{amssymb}
\usepackage{braket}
\expandafter\let\csname equation*\endcsname\relax 
\expandafter\let\csname endequation*\endcsname\relax
\usepackage{amsmath}

\usepackage[normalem]{ulem}
\usepackage{hyperref}
\usepackage{epstopdf}

\newcommand{\vect}[1]{\mathbf{#1}}

\def\be{\begin{equation}}
\def\ee{\end{equation}}
\def\bea{\begin{eqnarray}}
\def\eea{\end{eqnarray}}

\def\bi{\begin{itemize}}
\def\ei{\end{itemize}}
\def\ben{\begin{enumerate}}
\def\een{\end{enumerate}}

\begin{document}
\title{Controlled preparation of phases in two-dimensional time crystals}

\author{Arkadiusz Kuro\'s}
\email{arkadiusz.kuros@ujk.edu.pl}
\affiliation{Instytut Fizyki Teoretycznej, Uniwersytet Jagiello\'nski, ulica Profesora Stanis\l{}awa \L{}ojasiewicza 11, PL-30-348 Krak\'ow, Poland}
\affiliation{Institute of Physics, Jan Kochanowski University, Uniwersytecka 7, 25-406 Kielce, Poland}
\author{Rick~Mukherjee}
\email{r.mukherjee@imperial.ac.uk}
\affiliation{Blackett Laboratory, Imperial College London, SW7 2AZ, London, UK}
\author{Florian Mintert}
\affiliation{Blackett Laboratory, Imperial College London, SW7 2AZ, London, UK}
\author{Krzysztof Sacha}
\affiliation{Instytut Fizyki Teoretycznej, Uniwersytet Jagiello\'nski, ulica Profesora Stanis\l{}awa \L{}ojasiewicza 11, PL-30-348 Krak\'ow, Poland}

\date{\today}

\begin{abstract}
The study of phases is useful for understanding novel states of matter. One such state of matter are time crystals which constitute periodically driven interacting many-body systems that spontaneously break time translation symmetry. Time crystals with arbitrary periods (and dimensions) can be realized using the model of Bose-Einstein condensates bouncing on periodically-driven mirror(s). In this work, we identify the different phases that characterize the two-dimensional time crystal. By determining the optimal initial conditions and value of system parameters, we provide a practical route to realize a specific phase of the time crystal. These different phases can be mapped to the many-body states existing on a two-dimensional Hubbard lattice model, thereby opening up interesting opportunities for quantum simulation of many-body physics in time lattices.
\end{abstract}

\maketitle
\section{Introduction}
The ability to trap ultra-cold atomic gases and control their interactions with high precision has lead the way towards the realization of new phases of matter \cite{Lewenstein2007Adv,Goldman2014,Banuls2019review,Schafer2020}. These include the superfluid and Mott insulator phases of the Hubbard model \cite{Greiner}, topological states of matter \cite{Lin2, Goldman,Goldman2}, atoms with artificial gauge potentials \cite{Dalibard,Osterloh,Lin,Aidelsburger, banuls2020simulating}, supersolidity \cite{Natale, Bottcher} as well as many-body crystals \cite{Schauss,Waki, Britton}. Crystalline structures are examples of strongly correlated many-body systems generally resulting in spatially ordered configurations of  the constituent particles. In recent years, there has been considerable interest in studying a more unconventional type of crystals, namely discrete time crystals \cite{Sacha,ElseFTC,Khemani16,sacha2017time,sacha2020time}.

Discrete time crystals constitute periodically driven quantum many-body systems that spontaneously break discrete time translation symmetry and have been observed experimentally \cite{Monroe2017, choi2017observation, Pal2018, Rovny2018, smits2018observation}. So far these realizations of discrete time crystals have been restricted to systems that can be mapped to one dimensional (1D) models and the ratio of the period of their time evolution to the driving period was small ($\le$ 3). There are theoretical proposals that study discrete time crystals with larger periods {\cite{giergiel2018time,surace2019floquet, pizzi2019period, Kuro__2020, giergiel2020creating,Pizzi2021} as well as time lattices that map to higher dimensional lattice problems \cite{guo2013phase,  Giergiel,Liang2017, Giergiel2019,sacha2017time,guo2020condensed, sacha2020time, Giedrius}. The study of higher dimensional lattices in the context of discrete time crystals is appealing as it provides an additional degree of freedom to investigate the gradual time translation symmetry breaking processes and its connection with different discrete phases in the system. So far, such studies were done only in 1D models \cite{surace2019floquet, taheri2020all, yang2021dynamical,sakurai2021chimera}. 
\begin{figure}[t!] 	
	\centering              
	\includegraphics[width=1.0\columnwidth]{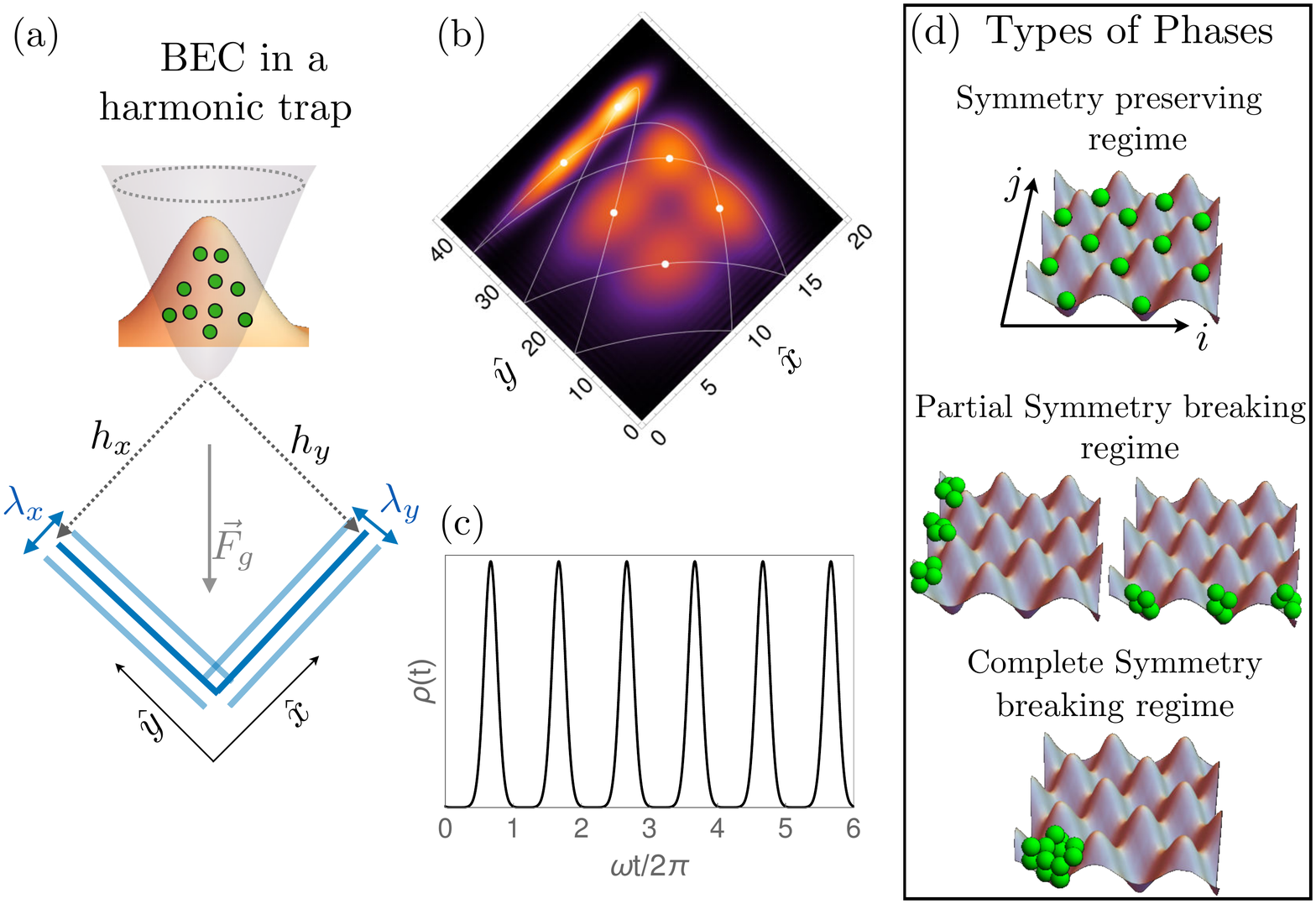} 
	\caption{(a) Ultra-cold atoms with attractive interactions are initially trapped in the lowest mode of a 2D harmonic trap. On its release, it falls under gravity (whose direction is indicated with $\vec{F}_g$) resulting in the bouncing of the atoms between the two harmonically oscillating orthogonal mirrors. Both mirrors oscillate with frequency $\omega$, but with individual amplitudes $\lambda_{x,y}$ respectively. In order to obtain stable dynamics, the initial conditions (position and momentum) need to be optimized which depend on heights $(h_x,h_y)$. (b) Density of non-interacting atoms bouncing between two oscillating mirrors at $t=2\pi/3 \omega$ and corresponding to a resonant Floquet state.  (c) The probability density for detecting a particle at fixed position $r=(16,37)$ in (b) at different times. (d) The system maps to an effective $s_x \times s_y$ Bose-Hubbard model. By tuning the strength of the attractive interactions, three different phases are identified. } 
	\label{Fig1_Setup}   
\end{figure}

Similar to the 1D case, one of the main challenges in realizing higher dimensional time lattices is to find appropriate initial conditions for the many-body quantum dynamics that follow periodic classical trajectories \cite{Kuro__2020}.  The system studied in the present paper consists of a Bose-Einstein condensate (BEC) bouncing on a pair of orthogonal atom mirrors that are periodically driven as shown schematically in Fig.~\ref{Fig1_Setup}(a). In the classical description, this system reveals nonlinear resonances and the motion for a single particle can be irregular. However, if the driving amplitude of the periodically driven mirror is small enough, there exist regular resonance islands in the phase space that are located around periodic orbits. For sufficiently large resonance islands, a quantum description is adopted where localized wave-packets [see Fig.~\ref{Fig1_Setup}(b)] that evolve along these periodic orbits form a basis of Wannier-like states of a tight-binding model that describes a 2D time lattice \cite{sacha2017time,Giergiel,Giergiel2019, guo2020condensed,sacha2020time}. Physically, this means that when we locate a detector close to the classical trajectory at fixed $\vect r=(x,y)$, the probability of its clicking will be periodic in time and reflects a cut of the 2D lattice as shown in Fig.~\ref{Fig1_Setup}(c). Choosing $\vect r$ close to different points on the classical trajectory, one can observe different cuts, which all together show a 2D crystalline structure in the time domain \cite{Giergiel,Giergiel2019}. In the presence of the attractive interactions between atoms, the system can break the time translation symmetry. The different crystalline phases reflect the degree to which the time translation symmetry is broken which is schematically depicted in Fig.~\ref{Fig1_Setup}(d).

In this work, we use statistical machine learning with Bayesian inference to find optimal conditions that realize robust discrete time crystals for higher dimensional lattices. The most general approach to find these suitable initial conditions would involve optimizing over $N$ particles for the many-body system as well as taking into account any possible noise that may occur in preparing the initial state. Such an optimization task is intractable even at the theoretical level. However we can simplify the optimization task by approximating the many-body wave function as a single quantum wave-packet thereby reducing the control parameters to a manageable number of six consisting of initial position, momentum and size of the wave-packet determined by the 2D harmonic trap in which the condensate is initially stored, see Fig.~\ref{Fig1_Setup}(a). This approximation is justified in the mean-field limit for the gas of bosonic atoms if the time required to prepare the initial state is much shorter than the overall dynamics of the discrete  time crystal. We report the existence of three distinct phases for the 2D time crystal where the gradual breaking of time translation symmetry can be achieved by either modulating the interaction strength between the atoms or by tuning the individual mirror amplitudes and frequency of the mirror oscillations. Moreover, one can selectively control the direction in which the time symmetry is broken which is reflected as selective filling of the lattice along a given direction in the  Bose-Hubbard picture as shown in the partial symmetry breaking regime in Fig.~\ref{Fig1_Setup}(d).

\section{Theory}
The system considered in this paper is a cloud of ultra-cold atoms bouncing on mirrors, but despite its many-body character, certain salient features of the model are best understood in the single particle picture \cite{Giergiel2019}. Thus, we first introduce the time-dependent model for the single particle which is naturally extended to incorporate the many-body Floquet Hamiltonian. The mapping of this system to the Bose-Hubbard model has been well studied \cite{giergiel2018time, Giergiel2019, guo2020condensed,sacha2020time} for which we provide a brief overview. Finally, we discuss the numerical method used to solve the many-body dynamics and the control techniques to obtain the optimal solutions.

\textit{Single-particle model:} The static Hamiltonian for the single-particle in 1D is classically integrable and the phase-space is completely foliated with periodic orbits on invariant tori. One finds that certain periodic orbits are localized inside the resonant islands when the mirror oscillations are on and provided  the amplitude of the oscillation is sufficiently small. In the quantum description, it implies that if a resonant island is large enough it can support one or more quantum states. 

Consider a single particle bouncing resonantly on a pair of oscillating orthogonal mirrors under the influence of gravity. We assume that the mirrors oscillate with the same frequency $\omega$. In the frame moving with the mirrors, the Hamiltonian of the system  is given by 
\be
H(t) =\sum_{\alpha=x,y} \left[\frac{p_{\alpha}^2}{2}+\alpha+\lambda_{\alpha} \alpha \cos(\omega t+\delta_{\alpha})\right],  \ \  \alpha \ge 0
\label{h}
\ee
where $\delta_{\alpha=x,y}$ and $\lambda_{\alpha=x,y}$ denote the phases and amplitudes of the mirror oscillations. In this work, all calculations are done in gravitational units but the gravitational acceleration is re-scaled by a factor $1/\sqrt{2}$. Since the mirrors are orthogonal, the single-particle dynamics separates into two independent motions along the $x$ and $y$ directions.

Using the Floquet theorem, one can obtain time-periodic eigenstates of the Floquet Hamiltonian $H(t)-i\partial_t$, which evolve with the driving period $T'=2\pi/\omega$ \cite{buchleitner2002non,Giergiel2019}. Defining $\Omega_x$ and $\Omega_y$ as frequencies for the unperturbed classical motion of the particle along the respective direction, we assume $s_x\Omega_x = s_y\Omega_y = \omega$,  where $s_{x,y}$ are integers. This is the condition for resonant driving of the 2D system. Shape of the resonant orbits in the 2D space depends on the ratio $\Omega_x/\Omega_y$ as well as on the relative phase of the mirrors $\delta=\delta_x-\delta_y$. If $s_{x,y}\gg1$, the quasi-energies corresponding to the resonant Floquet states form a band structure. Within the tight-binding approximation, we restrict the analysis to the first energy band of the single-particle Floquet Hamiltonian and construct 2D Wannier functions $W_{\vect i}(x,y,t)$ that are products of localized wave-packets $w_{i_x}(x,t)$ and $w_{i_y}(y,t)$ moving along the $x$ and $y$ directions with the periods $s_xT'$ and $s_yT'$, respectively \cite{Giergiel2019}. Here $\vect i=(i_x,i_y)$ is a double index with components in the range $i_x=1\hdots s_x$ and $i_y=1\hdots s_y$. The Wannier functions $W_{\vect i}(x,y,t)$ move in the 2D space along the classical resonant orbit with the period $T=s_xs_yT'$. 

\textit{Many-particle model:} The many-body Floquet Hamiltonian of ultra-cold bosonic atoms which are bouncing resonantly on the pair of oscillating  mirrors can be written in the form \cite{Sacha,sacha2020time,Wang2021two-mode},
\be
\hat{\cal H} =\frac{1}{T}\int\limits_0^{T}dt\int dxdy\;\hat\psi^\dagger\left[H(t) +\frac{g(t)}{2}\hat\psi^\dagger\hat\psi-i\partial_t\right]\hat\psi,
\label{mbfh}
\ee
where $H(t)$ is the single-particle Hamiltonian given in Eq.~(\ref{h}), $\hat \psi(x,y,t)$ is the bosonic field operator and $g(t) = g_0 f(t)$, where $g_0$ is the strength of the contact interactions between the atoms and $f(t)$ is an arbitrary periodic function with period $T$ which describes possible modulation of the strength of interactions between atoms in time. Expanding the  bosonic field operator in the Wannier basis, we get $\hat \psi(x,y,t)\approx\sum_{\vect i}\hat a_{\vect i}~W_{\vect i}(x,y,t)$ where $\hat a_{\vect i}$ are the bosonic annihilation operators. The description of a resonantly driven many-body system within the first energy band of the single particle system can be mapped to an effective tight-binding Hamiltonian
\be
\hat{\cal H} \approx -\frac{1}{2}\sum_{\langle \vect{i}, \vect{j}\rangle}J_{\vect i \vect j}\;\hat a_{\vect i}^\dagger\hat a_{\vect j}+\frac12\sum_{\vect i,\vect j}U_{\vect i\vect j}\;\hat a_{\vect i}^\dagger\hat a_{\vect j}^\dagger\hat a_{\vect j}\hat a_{\vect i}.
\label{bhh}
\ee
The above Hamiltonian is the Bose-Hubbard model in a time-periodic basis with the effective interaction coefficients 
\be
U_{\vect i \vect j}= (2-\delta_{\vect i \vect j})\frac{ N}{T}\int\limits_0^T g(t) dt\int dxdy~|W_{\vect i}|^2|W_{\vect j}|^2,
\label{inter}
\ee
and the tunneling amplitudes as
\be 
J_{\vect i \vect j}=-\frac{2}{T}\int^T_0 dt \int^{\infty}_0 dxdy~W^*_{\vect i}(t)\left[H(t)-i\partial_t\right]W_{\vect j}(t) .
\ee
Here we have assumed that the interaction energy per particle is smaller than the energy gap between the first and second quasi-energy bands of the single-particle system \cite{Giergiel2019} which limits the overall allowed strength of the interactions. The tunneling amplitudes $J_{\vect i \vect j}$ depend on the amplitudes and frequencies  of the mirrors' oscillations. In general, for attractive interactions, the ground state of the Hamiltonian (\ref{bhh}) within the mean-field approximation can be superposition of the Wannier states, i.e. 
\be
\psi(x,y, t)\approx \sum_{\vect i}a_{\vect i}W_{\vect i}(x,y,t)
\label{meanpsi}
\ee 
with complex amplitudes $a_{\vect i}$. Having derived the Bose-Hubbard model for the setup in the reduced Hilbert space (first energy band of the single particle system), it is often also useful to solve the mean-field BEC dynamics in the full Hilbert space in order to capture all the details of the dynamics.

\textit{BEC dynamics:} For a Bose-Einstein condensate, all $N$  atoms occupy the same single-particle state and the many-body wave-function factorizes as $ \phi(x_1,y_1,t)\phi(x_2,y_2,t)\hdots \phi(x_N,y_N,t)$  \cite{Pethick2002}. Within  the mean-field approximation, the single-particle state $\phi(x,y, t)$ satisfies the Gross–Pitaevskii (GP) equation \cite{Pethick2002}, 
\be
i\partial_t\phi(x,y, t)=\left[H(t)+g(t)N|\phi(x,y,t)|^2\right]\phi(x,y, t).
\label{gpfull}
\ee
Physically, the resonant dynamics corresponds to the coherent propagation of localized wave-packet along classical resonant orbit as shown in Fig.~\ref{Fig1_Setup}(b).  \\

\textit{Optimal control of the many-body dynamics:} The experimental realization of time crystals requires precise control over the initial conditions of the BEC dynamics. More specifically, the initial position and momentum of the quantum many-body wave-packet has to lie on the resonant classical trajectory. In order to determine the optimal initial conditions in the laboratory, we simulate experimental optimization by menas of the Bayesian optimization method since it tends to rapidly converge to optimal solutions for certain many-body problems \cite{Mukherjee2, Sauvage}. For sufficiently strong attractive interactions, the lowest energy state within the resonant Hilbert subspace can be described by a 2D wave-packet thereby reducing the control parameters for an $N$ particle quantum wave-packet to a tractable number of six, namely initial position  $(x_0,y_0)$, momentum $(p_{x_0}, p_{y_0})$ as well as the width of the wave-packet $(\sigma_{x_0},\sigma_{y_0})$  along each direction. The wave-packet widths are related to the harmonic trap frequencies along the relevant direction [see the schematic figure in Fig.~\ref{Fig1_Setup}(a)]. Furthermore, we can treat the initial cloud of atoms to be non-interacting similar to the single-particle problem. This approximation is valid provided the time to prepare the initial wave-packet is much smaller than the typical tunneling time between the wave-packets when the weak interactions influence the dynamics. In this limit, the Hamiltonian becomes separable along each dimension and we can perform optimization in either dimension independently. Without loss of generality, we focus on the $x$ direction and assume a Gaussian wave-packet as our initial state,
\be 
\phi(x,t=0)=\left(\frac{\tilde{\omega}_x}{\pi}\right)^{1/4}\exp{\left[-\frac{\tilde{\omega}_x(x-\tilde{x})^2}{2} -i\tilde{p}_{x}(x-\tilde{x})\right]} ,
\label{inopt}
\ee
where $\tilde{x}, \tilde{p}_x$ and $\tilde{\omega}_x$ are parameters to be determined ($\tilde{\omega}_x$ corresponds to the frequency of the harmonic trap where a BEC is initially prepared --- the width of the particle density in the trap equals $\sqrt{\tilde{\omega}_x}$). In order to simulate the experimental conditions, especially the noise in the initial conditions, we sample these parameters randomly from a uniform distribution: $\tilde{x} \in (x_0-\delta x_0, x_0+\delta x_0)$, $\tilde{p}_x \in (p_{x_0}-\delta p_{x_0}, p_{x_0}+\delta p_{x_0})$ and $\tilde{\omega}_x \in(\omega_{x}-\delta \omega_{x}, \omega_{x}+\delta \omega_{x})$. The  figure of merit
\be 
F_{M}  =D(2T)+D(3T), 
\label{fom}
\ee
is the sum of the overlaps 
\be 
D(t)=C_N\int dx |\phi(x,0)|^2 |\phi(x,t)|^2 ,
\ee
of the atomic densities, where $C_N=1/\int dx |\phi(x,0)|^4$ is the normalization constant. The merit of using $F_M$ rather than the squared overlap $|\braket{\phi(0)|\phi(t)}|^2$ as the figure of merit is that $F_M$, in contrast to the overlap, can easily be recovered experimentally from an average particle-number distribution. The overlap of densities is not sensitive to velocities two overlapping wavepackets pass each other. We have found that in order to mitigate this problem, one can use the sum of the density overlaps at two different moments of time, cf. Eq.~(\ref{fom}). The optimization was done using  the GPyOpt package \cite{gpyopt2016} and the choice of the acquisition function was expected improvement. For more details about Bayesian optimization, see Refs.\cite{snoek2012practical,frazier2018tutorial}.

\begin{figure}[t!] 	
\centering              
\includegraphics[width=1.0\columnwidth]{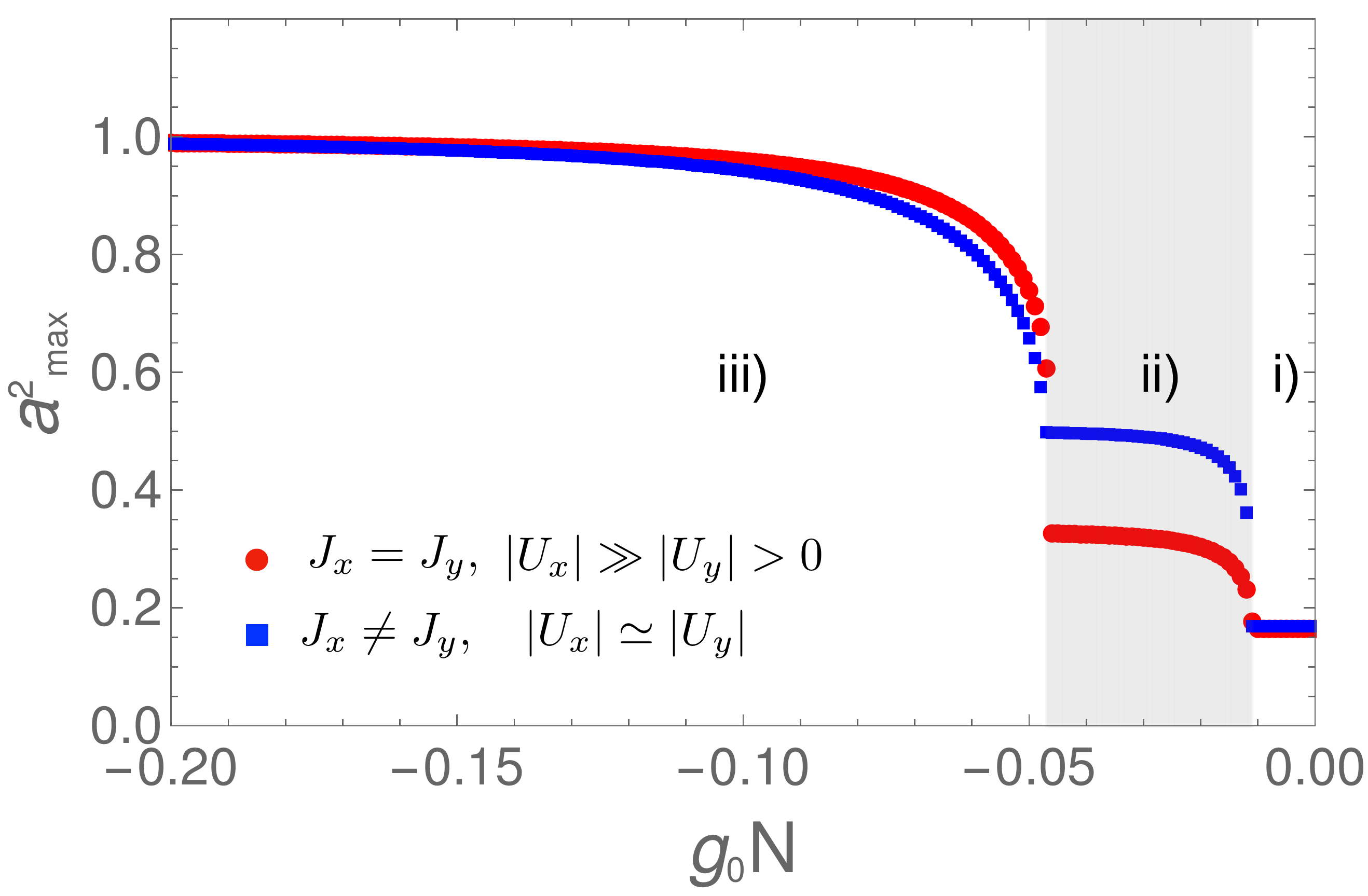} 
\caption{Three different regimes of the interaction strength (for the case of $s_x=2$ and $s_y=3$) characterized  by the parameter $a_{max}^2=max\{|a_{\vect i}|^2\}$, cf. Eq.~(\ref{meanpsi}). Different regions correspond to the ground state solutions of the Hamiltonian (\ref{bhh}) with time translation symmetry (i) being preserved in both lattice directions, (ii) broken only in one of the lattice direction and (iii) broken in both the lattice directions. Two different strategies have been used to break the symmetry. The first (depicted with red circles) corresponds to isotropic tunneling $J_x=J_y$ (where $J_x=J_{(i_x,i_y;i_x+1,i_y)}$ and $J_y=J_{(i_x,i_y;i_x,i_y+1)}$)  with significant nearest-neighbor interactions in one direction, $U_x\gg U_y$ (where $U_x=U_{(i_x,i_y;i_x+1,i_y)}$ and $U_y = U_{(i_x,i_y;i_x,i_y+1)}$). The other (depicted with blue squares) corresponds to anisotropic tunneling $J_x\ne J_y$. The former method results in $1/6<a_{max}^2<1/3$ while the latter gives $1/6<a_{max}^2<1/2$ in regime (ii). As a consequence of our choice of the system parameters, different symmetry regimes in both strategies coincide with each other --- in general they can be located at different ranges of $g_0N$. We use mirror amplitudes $\lambda_x=0.094$ and $\lambda_y = 0.03$, frequency $\omega = 1.1$ and relative phase of $\delta = 2 \pi /3$ to get isotropic tunnelling, $J_x=J_y =4.8 \times10^{-6}$. For anisotropic tunnelling rates, we used $\lambda_x=0.12$, $\lambda_y=0.09$, $\omega=1.4$ and $\delta = \pi/8$ giving $J_x=7.2 \times 10^{-4}$, $J_y=3.7\times 10^{-5}$.}
\label{Fig2_a_max}   
\end{figure}

\section{Results}
The main result of our work is the identification of the different phases that characterize the discrete time crystal in two dimensions. Typically the ground state of the Hamiltonian (\ref{bhh}) follows a discrete time translation symmetry which is spontaneously broken for sufficiently strong attractive interactions. However, when compared to the 1D time crystals, we find that the additional spatial degree provides more flexibility in breaking the time translation symmetry. For example, in this work, the time translation symmetry is also broken by selectively tuning the mirror oscillation amplitudes in either direction independently. Controlling the oscillation of the orthogonal pair of mirrors affects the tunneling amplitudes $J_{\vect i \vect j}$ in the Bose-Hubbard picture (\ref{bhh}). Although the different phases obtained for our periodically driven system can be understood by how strong the time translation symmetry is broken, it has a simple and elegant correspondence to the 2D Bose-Hubbard lattice as shown schematically in Fig.~\ref{Fig1_Setup}(d).
\begin{figure}[t!] 	
	\centering              
	\includegraphics[width=1.0\columnwidth]{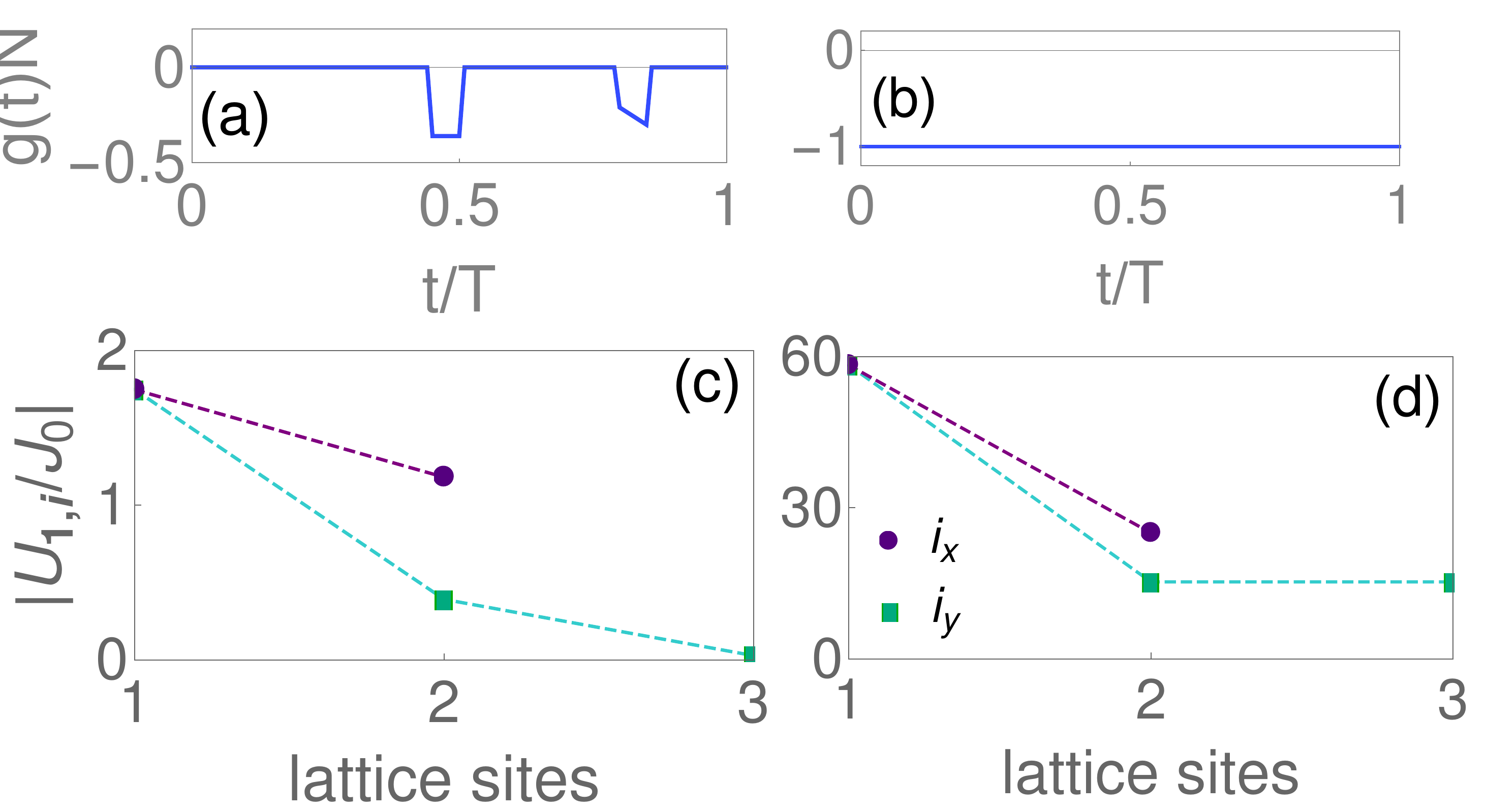} 
	\caption{(a-b) Modulation of the contact interactions between atoms as function of time. (c-d) Values of the interaction coefficients along the two lattice directions for the Bose-Hubbard model corresponding to (a-b) respectively. Small variation of the contact interactions is sufficient to generate non-negligible anisotropic interactions for nearest-neighbours.} 
	\label{Fig3_U}   
\end{figure}

The three relevant phases found in the model are shown in Fig.~\ref{Fig2_a_max}. When the time translation symmetry is preserved then the ground state is a uniform superposition of Wannier states in Eq.(\ref{meanpsi}) and the parameter $a^2_{max}=max\{|a_{\vect i}|^2\}$ takes the value $1/6$ (since $s_x=2$ and $s_y=3$)  which is represented by region (i) in Fig.~\ref{Fig2_a_max}. The scenario where the time translation symmetry is completely broken such that all $N$ atoms occupy a single site in the lattice model and ground state of the system can be described by a single Wannier state is represented as region (iii) in Fig.~\ref{Fig2_a_max}. The more interesting scenario is when the time translation symmetry is partially broken, in the sense that it is broken in one of the lattice direction but not the other which is given by (\ref{bhh}) and corresponds to region (ii) in Fig.~\ref{Fig2_a_max}. This occurs for higher dimensional ($d>1$) lattices, where the ground state of the Hamiltonian (\ref{bhh}) is a superposition of either $s_x$ or $s_y$ wave-packets depending on the direction in which the symmetry is broken. This phase is interesting because the corresponding ground state of the lattice model can have many different possibilities (in terms of lattice filling although a specific case has been schematically shown in Fig.~\ref{Fig1_Setup}(d)).

We recognize two different pathways of obtaining the phase (ii): (a) Directionally isotropic tunneling rates with anisotropic nearest-neighbour lattice interactions  shown with red circles in Fig.~\ref{Fig2_a_max}, and (b) directionally anisotropic tunneling rates shown with blue rectangles in Fig.~\ref{Fig2_a_max}. Thus depending on the which strategy is chosen, the  domain of the weakly interacting phase (ii) in Fig.~\ref{Fig2_a_max} is determined by the details of anisotropy of either the tunneling rates or lattice interaction strengths. As mentioned before, the tunneling amplitudes can be controlled by choosing specific values for the mirror oscillation amplitudes which can take a large range of values provided we satisfy the small amplitude approximation of the mirror oscillations \footnote{For this work, in gravitational units, we use mirror amplitudes $\lambda_x=0.094$ and $\lambda_y = 0.03$, frequency $\omega = 1.1$ and relative phase of $\delta = 2 \pi /3$ to get isotropic tunneling, $J_x=J_y =4.8 \times10^{-6}$. For anisotropic tunneling rates, we used $\lambda_x=0.12$, $\lambda_y=0.09$, $\omega=1.4$ and $\delta = \pi/8$ giving $J_x=7.2 \times 10^{-4}$, $J_y=3.7\times 10^{-5}$.}. Fig.~\ref{Fig3_U} shows that the interaction coefficients $|U_{\vect i \vect j}|$ in Eq.~\ref{inter} are controlled by modulating the scattering length for the contact interactions between the atoms at specific moments in time. These times correspond to the exact moments when Wannier states $W_{\textbf{i}}$ and $W_{\textbf{j}}$ pass each other. Thus, by mildly modulating $g(t)N$ over time, we get significant anisotropy in the nearest-neighbour interactions when compared to keeping the scattering length constant. Although the interaction modulations shown in Fig.~\ref{Fig3_U} are specific to the parameters chosen for Fig.~\ref{Fig2_a_max}, the protocol is completely generic for any set of parameters.
\begin{figure}[t!] 	
	\centering              
	\includegraphics[width=1.\columnwidth]{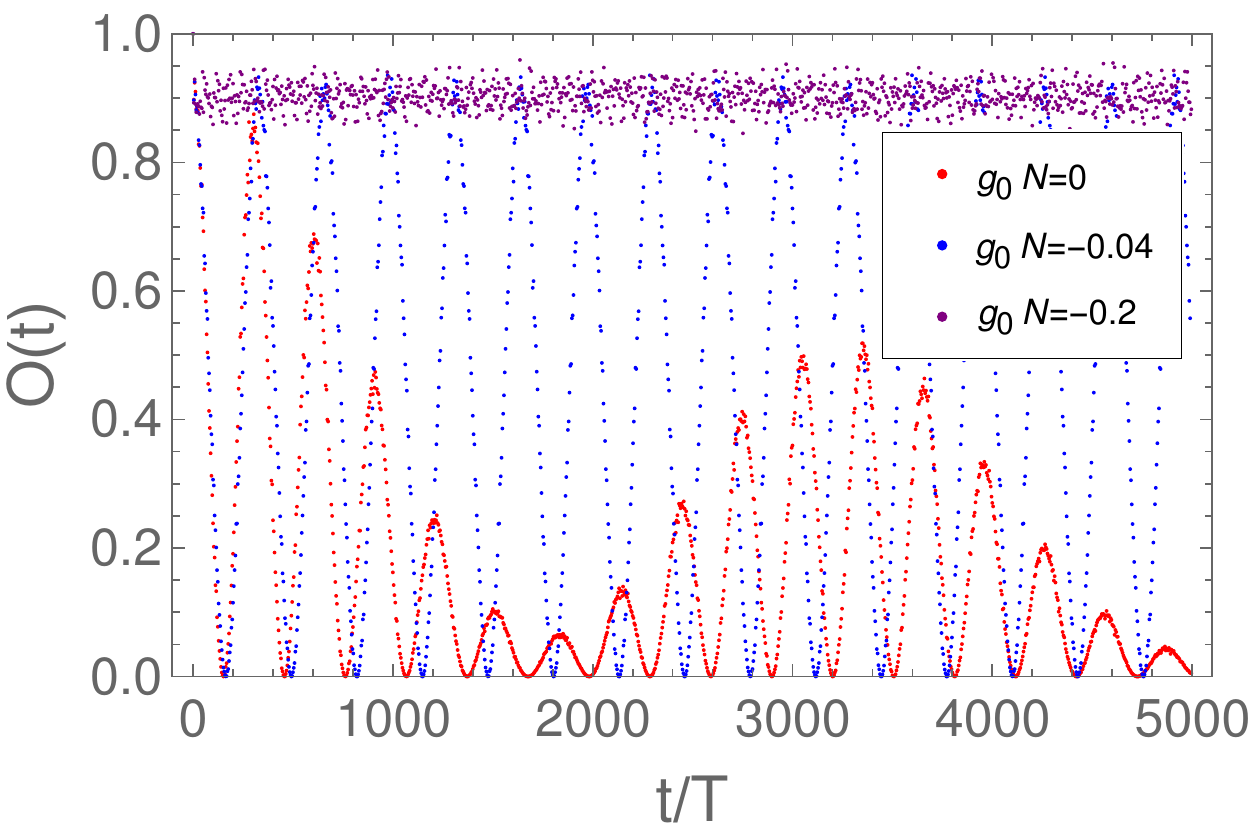} 
	\caption{Overlap between the instantaneous state and the initial state, $O(t)=|\int dx dy \phi^*(x,y,t) \phi(x,y,0) |^2$ for interaction strengths corresponding to the three different regimes: $g_0N=0$ (regime (i) in Fig.~\ref{Fig2_a_max}), $g_0N=-0.04$ [regime (ii)] and $g_0N=-0.2$ [regime (iii)]. The initial state has been chosen as $\phi(x,0)\phi(y,0)$ where $\phi$ defined in Eq.~(\ref{inopt}) is obtained by the optimization procedure.}
	\label{Fig4_fidelity}   
\end{figure}

Our next analysis is regarding the search of optimal initial conditions that can realize any of the discrete time crystal phases. We focus on the strongly interacting case for which we expect a discrete time crystal \cite{Sacha,giergiel2018time,Giergiel2019,giergiel2020creating} and use it to benchmark the required initial conditions for any arbitrary phase. Thus, $F_M(t)$ (see Eq.~\ref{fom}) is maximized with respect to the initial state parameters such that the initial state is periodically retrieved at long times in integer multiples of $T=s_xs_yT'$. $F_M(t)$ is evaluated from $\phi(x,t)$, which is obtained by numerically solving the Gross–Pitaevskii Eq.~\ref{gpfull} using the split-step fast Fourier transform method with step-sizes $dx=0.002x_0$ and $dt=0.001T$. The Bayesian optimization provided the optimal parameters for initial wave-packet defined in Eq.~(\ref{inopt}) \footnote{In gravitational units, $\tilde{x}\in [0.98, 1.02]x_0$, $\tilde{p}_{x}\in [0.98, 1.02] p_{x_0}$ and $\tilde{\omega}_{x} \in [0.98, 1.02] \omega_{x}$, where $x_0 = 9.82$, $p_{x_0}= 0.28$ and $\omega_{x} =0.68$. Similarly for the $y$-direction, we have $y_0 = 22.41$, $p_{y_0} =-0.42$ and $\omega_{y}  = 0.53$ with the same range.}, which were obtained with 20 initial points, 100 iterations and averaged over 10 different noise realizations. However, since optimization of the initial state is performed in the short-time scale limit, it is independent of the interaction strength. In order to test how the optimized state evolves in the presence of the interactions we have integrated the GP equation. Using the same parameters as in the anisotropic tunneling case in Fig.~\ref{Fig2_a_max}, the results of these calculations are depicted in Fig.~\ref{Fig4_fidelity}  which shows the overlap of the optimized initial state with its time evolution $\phi(x,y,t)$ at long times for different interactions strengths. As expected, one can see that stable BEC dynamics in 2D is possible if the interactions are sufficiently strong and the system performs periodic evolution for a long time. The presented results are based on the mean-field approach. However, since the relation of the interaction energy per particle to the energy gap between the bands of the resonant quasi-energies is similar as in Refs.~\cite{Kuro__2020,wang2021many,Wang2021two-mode} where quantum many-body effects are analyzed, we also expect that the beyond mean-field approach will not show any signature of heating of the system by the periodic drive.

\begin{figure}[t!] 	
\centering              
\includegraphics[width=1.0\columnwidth]{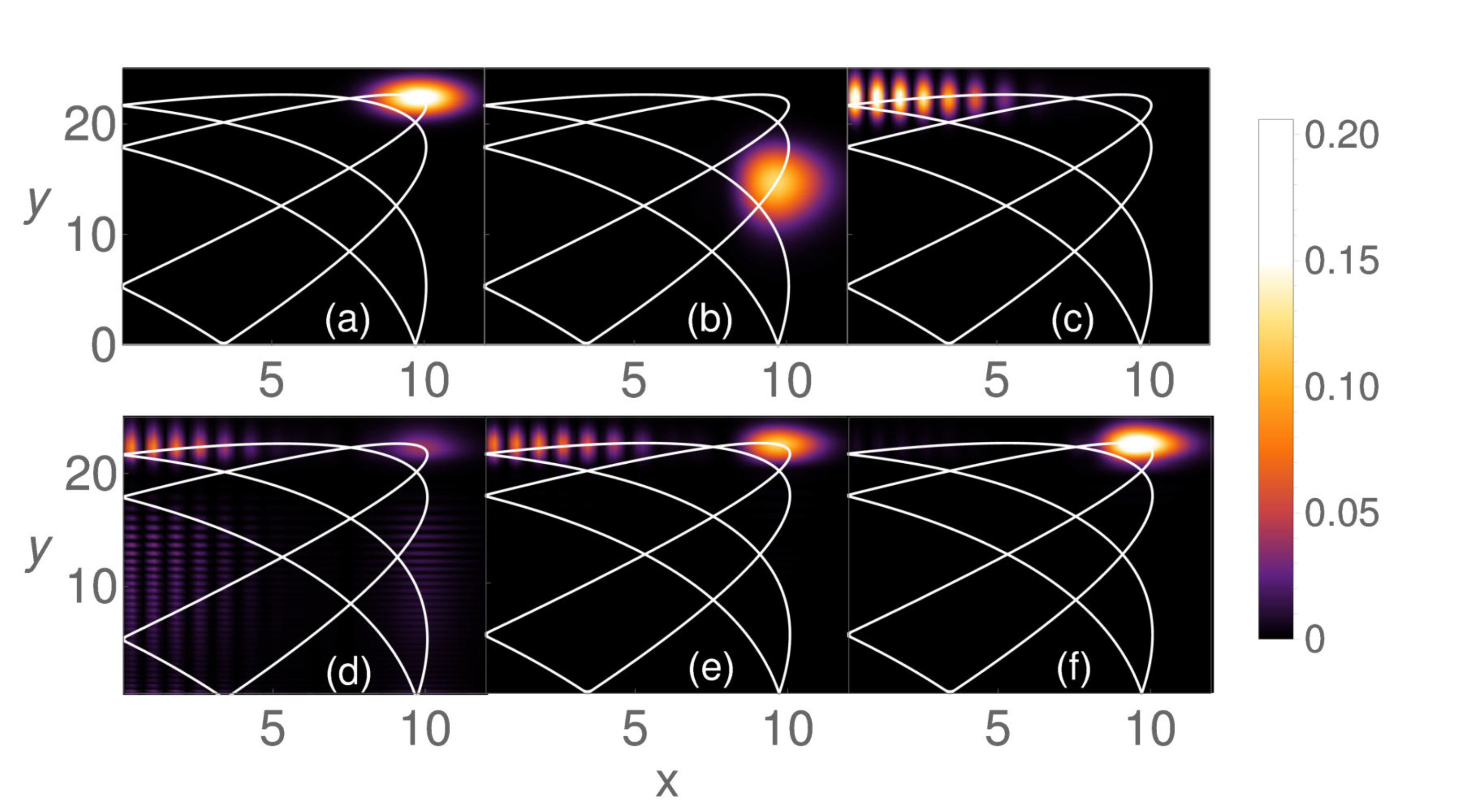} 
\caption{Density plot of BEC dynamics obtained with optimized parameters for the initial state. The white curves represent classical trajectories. (a-c) correspond to short time dynamics, which is the same for any $g_0$, for three different times (a) $t=0$, (b) $t=T/3$ and (c) $t =T/2$. (d-f) correspond to long-time dynamics for different interactions, (d) $g_0N=0$, (e) $g_0N=-0.04$ and (f) $g_0N=-0.2$ at fixed time $t=700T$. }
\label{Fig5_2Ddensity}   
\end{figure}

Results of the integration of the GP equation in the 2D space show also that the description of the system indeed reduces to the resonant Hilbert subspace spanned the $s_x s_y$ Wannier-like wave-packets, cf. Eq.~(\ref{meanpsi}). In Fig.~\ref{Fig5_2Ddensity} we present time evolution of the density of atoms starting with the optimized initial wave-packet for different interaction strengths. Within the single period $T$, we find that the  wave-packet is moving along the classical resonant orbit (white curve in Fig.~\ref{Fig5_2Ddensity}) and only  interference fringes are observed when it hits a mirror [cf. Fig.~\ref{Fig5_2Ddensity} (c)]. It should be noted that the short time dynamics ($t\le T$) is almost independent of $g_0$, see Fig.~\ref{Fig5_2Ddensity}(a-c). The reason for this is the interactions are very weak and can only modify tunneling process of atoms between different wave-packets which takes place at much longer time, i.e. $t\approx 1/J \gg T$. At long time scales, the interactions play a crucial role in the dynamics which is clearly visible in Fig.~\ref{Fig5_2Ddensity}(d-e). For weak interactions (almost non-interacting system), the localized wave-packet starts spreading into the six Wannier states, while for sufficiently strong interactions ($|g_0N|>0.1$) the wave-packet remains localized indicating that one particular Wannier state is dominant. This is consistent with Fig.~\ref{Fig2_a_max}. The partial symmetry breaking regime corresponds to suppression of tunneling along one of the directions in the lattice described by the Bose-Hubbard model (\ref{bhh}), see Fig.~\ref{Fig1_Setup}(d).

\section{Concluding Remarks}
In this work, we characterize the different phases realizable in a time crystal that map to a 2D lattice model. We find that one of the benefits of time crystals with properties of higher dimensional systems is the higher degree of freedom in controlling the system and preparing it in a certain phase. This is especially reflected in the scenario where the gradual breaking of time translation symmetry in either lattice directions is achieved by  selectively varying the system parameters. Optimal control was used not only in realizing the time crystals but also to observe signatures of the different phases.

The different phases correspond to the state which evolves with the period $T'$, $s_x T'$ or $s_y T'$ and $(s_x \times s_y)T'$ respectively. In order to distinguish the partial symmetry breaking regime, one should prepare the initial state as a superposition of $s_x$ or $s_y$ localized wave-packets moving with different velocities and with a specific relative phase between them, which is experimentally challenging. Alternatively it is much easier to prepare a single localized wave-packet and monitor its evolution along the resonant orbit, but this implies that one can observe only signatures of different phases from the particle density using time-of flight measurements. Although our analysis was done for $s_x=2$ and $s_y=3$ number of resonances along each direction, it is expected to be valid for higher resonances which is more suitable for experiments \cite{giergiel2018time,giergiel2020creating}. Already there exist experimental realizations similar to the setup described in this work \cite{Steane95, Roach1995,Sidorov1996,Westbrook1998,Lau1999,Bongs1999,Sidorov2002,Fiutowski2013,Kawalec2014}. The modulation of the interaction is routinely done by changing the  s-wave scattering length using Feshbach resonance mechanism \cite{Hadzibabic, Amo, Weiler}. Typical values of the tunneling rates for the lattice would be in the order of tens of Hz while the interaction coefficients would range from  tens of Hz to tens of kHz \cite{giergiel2018time}. Although our numerical results suggest stable BEC dynamics, further investigation of the effects of quantum heating would be useful. 
 
 The discrete time crystals with properties of higher dimensional lattice systems are in general appealing for simulating novel physics in condensed matter physics \cite{lustig2018topological, liang2018floquet, Giergiel_2019,Giedrius, giergiel2021inseparable} , most of which are yet to be realized in real experiments. The use of Bayesian optimizers for real experiments can be useful as it performs better with noisy control landscape \cite{Mukherjee_2020}. The ability for the Bayesian optimizer to find optimal initial conditions for BEC dynamics can have more general applications apart from constructing time crystals. For example, it can be used to efficiently transfer BEC from an initial harmonic trap into a desired state with high fidelity \cite{TORRONTEGUI2013117}. The desired state can be a particular band in an optical lattice \cite{Zhou}, a specific initial state needed for coherent BEC dynamics under the influence of gravity \cite{Bongs, Robert_de_Saint_Vincent_2010} or an initial set of conditions required for observing stable soliton dynamics \cite{Strecker, Nguyen}. Examples of controlled continuous loading of a BEC have relevant implications for an atomic laser \cite{Mewes} and in reducing two- and three-body losses, thereby enhancing the lifetime of typical BEC experiments\cite{Santos}.
 
\vspace{0.2cm}
\section*{Acknowledgments}
This work was supported by the National Science Centre, Poland via Projects QuantERA No. 2017/25/Z/ST2/03027 (AK) and No. 2018/31/B/ST2/00349 (KS). The authors acknowledge the funding from the QuantERA ERANET Cofund in Quantum Technologies implemented within the European Union’s Horizon 2020 Programme under the project Theory-Blind Quantum Control TheBlinQC and from EPSRC under the grant EP/R044082/1 (RM and FM).

\bibliography{refs.bib}

\begin{thebibliography}{77}
\expandafter\ifx\csname natexlab\endcsname\relax\def\natexlab#1{#1}\fi
\expandafter\ifx\csname bibnamefont\endcsname\relax
  \def\bibnamefont#1{#1}\fi
\expandafter\ifx\csname bibfnamefont\endcsname\relax
  \def\bibfnamefont#1{#1}\fi
\expandafter\ifx\csname citenamefont\endcsname\relax
  \def\citenamefont#1{#1}\fi
\expandafter\ifx\csname url\endcsname\relax
  \def\url#1{\texttt{#1}}\fi
\expandafter\ifx\csname urlprefix\endcsname\relax\def\urlprefix{URL }\fi
\providecommand{\bibinfo}[2]{#2}
\providecommand{\eprint}[2][]{\url{#2}}

\bibitem[{\citenamefont{Lewenstein et~al.}(2007)\citenamefont{Lewenstein,
  Sanpera, Ahufinger, Damski, Sen(De), and Sen}}]{Lewenstein2007Adv}
\bibinfo{author}{\bibfnamefont{M.}~\bibnamefont{Lewenstein}},
  \bibinfo{author}{\bibfnamefont{A.}~\bibnamefont{Sanpera}},
  \bibinfo{author}{\bibfnamefont{V.}~\bibnamefont{Ahufinger}},
  \bibinfo{author}{\bibfnamefont{B.}~\bibnamefont{Damski}},
  \bibinfo{author}{\bibfnamefont{A.}~\bibnamefont{Sen(De)}}, \bibnamefont{and}
  \bibinfo{author}{\bibfnamefont{U.}~\bibnamefont{Sen}},
  \bibinfo{journal}{Advances in Physics} \textbf{\bibinfo{volume}{56}},
  \bibinfo{pages}{243} (\bibinfo{year}{2007}),
  \eprint{https://doi.org/10.1080/00018730701223200},
  \urlprefix\url{https://doi.org/10.1080/00018730701223200}.

\bibitem[{\citenamefont{Goldman et~al.}(2014)\citenamefont{Goldman,
  Juzeli{\={u}}nas, Öhberg, and Spielman}}]{Goldman2014}
\bibinfo{author}{\bibfnamefont{N.}~\bibnamefont{Goldman}},
  \bibinfo{author}{\bibfnamefont{G.}~\bibnamefont{Juzeli{\={u}}nas}},
  \bibinfo{author}{\bibfnamefont{P.}~\bibnamefont{Öhberg}}, \bibnamefont{and}
  \bibinfo{author}{\bibfnamefont{I.~B.} \bibnamefont{Spielman}},
  \bibinfo{journal}{Reports on Progress in Physics}
  \textbf{\bibinfo{volume}{77}}, \bibinfo{pages}{126401}
  (\bibinfo{year}{2014}),
  \urlprefix\url{https://doi.org/10.1088/0034-4885/77/12/126401}.

\bibitem[{\citenamefont{{Ba{\~n}uls} et~al.}(2020)\citenamefont{{Ba{\~n}uls},
  {Blatt}, {Catani}, {Celi}, {Cirac}, {Dalmonte}, {Fallani}, {Jansen},
  {Lewenstein}, {Montangero} et~al.}}]{Banuls2019review}
\bibinfo{author}{\bibfnamefont{M.~C.} \bibnamefont{{Ba{\~n}uls}}},
  \bibinfo{author}{\bibfnamefont{R.}~\bibnamefont{{Blatt}}},
  \bibinfo{author}{\bibfnamefont{J.}~\bibnamefont{{Catani}}},
  \bibinfo{author}{\bibfnamefont{A.}~\bibnamefont{{Celi}}},
  \bibinfo{author}{\bibfnamefont{J.~I.} \bibnamefont{{Cirac}}},
  \bibinfo{author}{\bibfnamefont{M.}~\bibnamefont{{Dalmonte}}},
  \bibinfo{author}{\bibfnamefont{L.}~\bibnamefont{{Fallani}}},
  \bibinfo{author}{\bibfnamefont{K.}~\bibnamefont{{Jansen}}},
  \bibinfo{author}{\bibfnamefont{M.}~\bibnamefont{{Lewenstein}}},
  \bibinfo{author}{\bibfnamefont{S.}~\bibnamefont{{Montangero}}},
  \bibnamefont{et~al.}, \bibinfo{journal}{European Physical Journal D}
  \textbf{\bibinfo{volume}{74}}, \bibinfo{eid}{165} (\bibinfo{year}{2020}),
  \eprint{1911.00003}.

\bibitem[{\citenamefont{Sch{\"a}fer et~al.}(2020)\citenamefont{Sch{\"a}fer,
  Fukuhara, Sugawa, Takasu, and Takahashi}}]{Schafer2020}
\bibinfo{author}{\bibfnamefont{F.}~\bibnamefont{Sch{\"a}fer}},
  \bibinfo{author}{\bibfnamefont{T.}~\bibnamefont{Fukuhara}},
  \bibinfo{author}{\bibfnamefont{S.}~\bibnamefont{Sugawa}},
  \bibinfo{author}{\bibfnamefont{Y.}~\bibnamefont{Takasu}}, \bibnamefont{and}
  \bibinfo{author}{\bibfnamefont{Y.}~\bibnamefont{Takahashi}},
  \bibinfo{journal}{Nature Reviews Physics} \textbf{\bibinfo{volume}{2}},
  \bibinfo{pages}{411} (\bibinfo{year}{2020}).

\bibitem[{\citenamefont{Greiner et~al.}(2002)\citenamefont{Greiner, Mandel,
  Esslinger, H{\"a}nsch, and Bloch}}]{Greiner}
\bibinfo{author}{\bibfnamefont{M.}~\bibnamefont{Greiner}},
  \bibinfo{author}{\bibfnamefont{O.}~\bibnamefont{Mandel}},
  \bibinfo{author}{\bibfnamefont{T.}~\bibnamefont{Esslinger}},
  \bibinfo{author}{\bibfnamefont{T.~W.} \bibnamefont{H{\"a}nsch}},
  \bibnamefont{and} \bibinfo{author}{\bibfnamefont{I.}~\bibnamefont{Bloch}},
  \bibinfo{journal}{Nature} \textbf{\bibinfo{volume}{415}}, \bibinfo{pages}{39}
  (\bibinfo{year}{2002}).

\bibitem[{\citenamefont{Lin et~al.}(2011)\citenamefont{Lin,
  Jim{\'e}nez-Garc{\'\i}a, and Spielman}}]{Lin2}
\bibinfo{author}{\bibfnamefont{Y.~J.} \bibnamefont{Lin}},
  \bibinfo{author}{\bibfnamefont{K.}~\bibnamefont{Jim{\'e}nez-Garc{\'\i}a}},
  \bibnamefont{and} \bibinfo{author}{\bibfnamefont{I.~B.}
  \bibnamefont{Spielman}}, \bibinfo{journal}{Nature}
  \textbf{\bibinfo{volume}{471}}, \bibinfo{pages}{83} (\bibinfo{year}{2011}).

\bibitem[{\citenamefont{Goldman et~al.}(2009)\citenamefont{Goldman, Kubasiak,
  Bermudez, Gaspard, Lewenstein, and Martin-Delgado}}]{Goldman}
\bibinfo{author}{\bibfnamefont{N.}~\bibnamefont{Goldman}},
  \bibinfo{author}{\bibfnamefont{A.}~\bibnamefont{Kubasiak}},
  \bibinfo{author}{\bibfnamefont{A.}~\bibnamefont{Bermudez}},
  \bibinfo{author}{\bibfnamefont{P.}~\bibnamefont{Gaspard}},
  \bibinfo{author}{\bibfnamefont{M.}~\bibnamefont{Lewenstein}},
  \bibnamefont{and} \bibinfo{author}{\bibfnamefont{M.~A.}
  \bibnamefont{Martin-Delgado}}, \bibinfo{journal}{Phys. Rev. Lett.}
  \textbf{\bibinfo{volume}{103}}, \bibinfo{pages}{035301}
  (\bibinfo{year}{2009}),
  \urlprefix\url{https://link.aps.org/doi/10.1103/PhysRevLett.103.035301}.

\bibitem[{\citenamefont{Goldman et~al.}(2016)\citenamefont{Goldman, Budich, and
  Zoller}}]{Goldman2}
\bibinfo{author}{\bibfnamefont{N.}~\bibnamefont{Goldman}},
  \bibinfo{author}{\bibfnamefont{J.~C.} \bibnamefont{Budich}},
  \bibnamefont{and} \bibinfo{author}{\bibfnamefont{P.}~\bibnamefont{Zoller}},
  \bibinfo{journal}{Nature Physics} \textbf{\bibinfo{volume}{12}},
  \bibinfo{pages}{639} (\bibinfo{year}{2016}).

\bibitem[{\citenamefont{Dalibard et~al.}(2011)\citenamefont{Dalibard, Gerbier,
  Juzeli\ifmmode~\bar{u}\else \={u}\fi{}nas, and \"Ohberg}}]{Dalibard}
\bibinfo{author}{\bibfnamefont{J.}~\bibnamefont{Dalibard}},
  \bibinfo{author}{\bibfnamefont{F.}~\bibnamefont{Gerbier}},
  \bibinfo{author}{\bibfnamefont{G.}~\bibnamefont{Juzeli\ifmmode~\bar{u}\else
  \={u}\fi{}nas}}, \bibnamefont{and}
  \bibinfo{author}{\bibfnamefont{P.}~\bibnamefont{\"Ohberg}},
  \bibinfo{journal}{Rev. Mod. Phys.} \textbf{\bibinfo{volume}{83}},
  \bibinfo{pages}{1523} (\bibinfo{year}{2011}),
  \urlprefix\url{https://link.aps.org/doi/10.1103/RevModPhys.83.1523}.

\bibitem[{\citenamefont{Osterloh et~al.}(2005)\citenamefont{Osterloh, Baig,
  Santos, Zoller, and Lewenstein}}]{Osterloh}
\bibinfo{author}{\bibfnamefont{K.}~\bibnamefont{Osterloh}},
  \bibinfo{author}{\bibfnamefont{M.}~\bibnamefont{Baig}},
  \bibinfo{author}{\bibfnamefont{L.}~\bibnamefont{Santos}},
  \bibinfo{author}{\bibfnamefont{P.}~\bibnamefont{Zoller}}, \bibnamefont{and}
  \bibinfo{author}{\bibfnamefont{M.}~\bibnamefont{Lewenstein}},
  \bibinfo{journal}{Phys. Rev. Lett.} \textbf{\bibinfo{volume}{95}},
  \bibinfo{pages}{010403} (\bibinfo{year}{2005}),
  \urlprefix\url{https://link.aps.org/doi/10.1103/PhysRevLett.95.010403}.

\bibitem[{\citenamefont{Lin et~al.}(2009)\citenamefont{Lin, Compton, Perry,
  Phillips, Porto, and Spielman}}]{Lin}
\bibinfo{author}{\bibfnamefont{Y.-J.} \bibnamefont{Lin}},
  \bibinfo{author}{\bibfnamefont{R.~L.} \bibnamefont{Compton}},
  \bibinfo{author}{\bibfnamefont{A.~R.} \bibnamefont{Perry}},
  \bibinfo{author}{\bibfnamefont{W.~D.} \bibnamefont{Phillips}},
  \bibinfo{author}{\bibfnamefont{J.~V.} \bibnamefont{Porto}}, \bibnamefont{and}
  \bibinfo{author}{\bibfnamefont{I.~B.} \bibnamefont{Spielman}},
  \bibinfo{journal}{Phys. Rev. Lett.} \textbf{\bibinfo{volume}{102}},
  \bibinfo{pages}{130401} (\bibinfo{year}{2009}),
  \urlprefix\url{https://link.aps.org/doi/10.1103/PhysRevLett.102.130401}.

\bibitem[{\citenamefont{Aidelsburger et~al.}(2015)\citenamefont{Aidelsburger,
  Lohse, Schweizer, Atala, Barreiro, Nascimb{\`e}ne, Cooper, Bloch, and
  Goldman}}]{Aidelsburger}
\bibinfo{author}{\bibfnamefont{M.}~\bibnamefont{Aidelsburger}},
  \bibinfo{author}{\bibfnamefont{M.}~\bibnamefont{Lohse}},
  \bibinfo{author}{\bibfnamefont{C.}~\bibnamefont{Schweizer}},
  \bibinfo{author}{\bibfnamefont{M.}~\bibnamefont{Atala}},
  \bibinfo{author}{\bibfnamefont{J.~T.} \bibnamefont{Barreiro}},
  \bibinfo{author}{\bibfnamefont{S.}~\bibnamefont{Nascimb{\`e}ne}},
  \bibinfo{author}{\bibfnamefont{N.~R.} \bibnamefont{Cooper}},
  \bibinfo{author}{\bibfnamefont{I.}~\bibnamefont{Bloch}}, \bibnamefont{and}
  \bibinfo{author}{\bibfnamefont{N.}~\bibnamefont{Goldman}},
  \bibinfo{journal}{Nature Physics} \textbf{\bibinfo{volume}{11}},
  \bibinfo{pages}{162} (\bibinfo{year}{2015}).

\bibitem[{\citenamefont{Banuls et~al.}(2020)\citenamefont{Banuls, Blatt,
  Catani, Celi, Cirac, Dalmonte, Fallani, Jansen, Lewenstein, Montangero
  et~al.}}]{banuls2020simulating}
\bibinfo{author}{\bibfnamefont{M.~C.} \bibnamefont{Banuls}},
  \bibinfo{author}{\bibfnamefont{R.}~\bibnamefont{Blatt}},
  \bibinfo{author}{\bibfnamefont{J.}~\bibnamefont{Catani}},
  \bibinfo{author}{\bibfnamefont{A.}~\bibnamefont{Celi}},
  \bibinfo{author}{\bibfnamefont{J.~I.} \bibnamefont{Cirac}},
  \bibinfo{author}{\bibfnamefont{M.}~\bibnamefont{Dalmonte}},
  \bibinfo{author}{\bibfnamefont{L.}~\bibnamefont{Fallani}},
  \bibinfo{author}{\bibfnamefont{K.}~\bibnamefont{Jansen}},
  \bibinfo{author}{\bibfnamefont{M.}~\bibnamefont{Lewenstein}},
  \bibinfo{author}{\bibfnamefont{S.}~\bibnamefont{Montangero}},
  \bibnamefont{et~al.}, \bibinfo{journal}{The European physical journal D}
  \textbf{\bibinfo{volume}{74}}, \bibinfo{pages}{1} (\bibinfo{year}{2020}).

\bibitem[{\citenamefont{Natale et~al.}(2019)\citenamefont{Natale, van Bijnen,
  Patscheider, Petter, Mark, Chomaz, and Ferlaino}}]{Natale}
\bibinfo{author}{\bibfnamefont{G.}~\bibnamefont{Natale}},
  \bibinfo{author}{\bibfnamefont{R.~M.~W.} \bibnamefont{van Bijnen}},
  \bibinfo{author}{\bibfnamefont{A.}~\bibnamefont{Patscheider}},
  \bibinfo{author}{\bibfnamefont{D.}~\bibnamefont{Petter}},
  \bibinfo{author}{\bibfnamefont{M.~J.} \bibnamefont{Mark}},
  \bibinfo{author}{\bibfnamefont{L.}~\bibnamefont{Chomaz}}, \bibnamefont{and}
  \bibinfo{author}{\bibfnamefont{F.}~\bibnamefont{Ferlaino}},
  \bibinfo{journal}{Phys. Rev. Lett.} \textbf{\bibinfo{volume}{123}},
  \bibinfo{pages}{050402} (\bibinfo{year}{2019}),
  \urlprefix\url{https://link.aps.org/doi/10.1103/PhysRevLett.123.050402}.

\bibitem[{\citenamefont{B\"ottcher et~al.}(2019)\citenamefont{B\"ottcher,
  Schmidt, Wenzel, Hertkorn, Guo, Langen, and Pfau}}]{Bottcher}
\bibinfo{author}{\bibfnamefont{F.}~\bibnamefont{B\"ottcher}},
  \bibinfo{author}{\bibfnamefont{J.-N.} \bibnamefont{Schmidt}},
  \bibinfo{author}{\bibfnamefont{M.}~\bibnamefont{Wenzel}},
  \bibinfo{author}{\bibfnamefont{J.}~\bibnamefont{Hertkorn}},
  \bibinfo{author}{\bibfnamefont{M.}~\bibnamefont{Guo}},
  \bibinfo{author}{\bibfnamefont{T.}~\bibnamefont{Langen}}, \bibnamefont{and}
  \bibinfo{author}{\bibfnamefont{T.}~\bibnamefont{Pfau}},
  \bibinfo{journal}{Phys. Rev. X} \textbf{\bibinfo{volume}{9}},
  \bibinfo{pages}{011051} (\bibinfo{year}{2019}),
  \urlprefix\url{https://link.aps.org/doi/10.1103/PhysRevX.9.011051}.

\bibitem[{\citenamefont{Schau{\ss} et~al.}(2012)\citenamefont{Schau{\ss},
  Cheneau, Endres, Fukuhara, Hild, Omran, Pohl, Gross, Kuhr, and
  Bloch}}]{Schauss}
\bibinfo{author}{\bibfnamefont{P.}~\bibnamefont{Schau{\ss}}},
  \bibinfo{author}{\bibfnamefont{M.}~\bibnamefont{Cheneau}},
  \bibinfo{author}{\bibfnamefont{M.}~\bibnamefont{Endres}},
  \bibinfo{author}{\bibfnamefont{T.}~\bibnamefont{Fukuhara}},
  \bibinfo{author}{\bibfnamefont{S.}~\bibnamefont{Hild}},
  \bibinfo{author}{\bibfnamefont{A.}~\bibnamefont{Omran}},
  \bibinfo{author}{\bibfnamefont{T.}~\bibnamefont{Pohl}},
  \bibinfo{author}{\bibfnamefont{C.}~\bibnamefont{Gross}},
  \bibinfo{author}{\bibfnamefont{S.}~\bibnamefont{Kuhr}}, \bibnamefont{and}
  \bibinfo{author}{\bibfnamefont{I.}~\bibnamefont{Bloch}},
  \bibinfo{journal}{Nature} \textbf{\bibinfo{volume}{491}}, \bibinfo{pages}{87}
  (\bibinfo{year}{2012}).

\bibitem[{\citenamefont{Waki et~al.}(1992)\citenamefont{Waki, Kassner, Birkl,
  and Walther}}]{Waki}
\bibinfo{author}{\bibfnamefont{I.}~\bibnamefont{Waki}},
  \bibinfo{author}{\bibfnamefont{S.}~\bibnamefont{Kassner}},
  \bibinfo{author}{\bibfnamefont{G.}~\bibnamefont{Birkl}}, \bibnamefont{and}
  \bibinfo{author}{\bibfnamefont{H.}~\bibnamefont{Walther}},
  \bibinfo{journal}{Phys. Rev. Lett.} \textbf{\bibinfo{volume}{68}},
  \bibinfo{pages}{2007} (\bibinfo{year}{1992}),
  \urlprefix\url{https://link.aps.org/doi/10.1103/PhysRevLett.68.2007}.

\bibitem[{\citenamefont{Britton et~al.}(2012)\citenamefont{Britton, Sawyer,
  Keith, Wang, Freericks, Uys, Biercuk, and Bollinger}}]{Britton}
\bibinfo{author}{\bibfnamefont{J.~W.} \bibnamefont{Britton}},
  \bibinfo{author}{\bibfnamefont{B.~C.} \bibnamefont{Sawyer}},
  \bibinfo{author}{\bibfnamefont{A.~C.} \bibnamefont{Keith}},
  \bibinfo{author}{\bibfnamefont{C.~C.~J.} \bibnamefont{Wang}},
  \bibinfo{author}{\bibfnamefont{J.~K.} \bibnamefont{Freericks}},
  \bibinfo{author}{\bibfnamefont{H.}~\bibnamefont{Uys}},
  \bibinfo{author}{\bibfnamefont{M.~J.} \bibnamefont{Biercuk}},
  \bibnamefont{and} \bibinfo{author}{\bibfnamefont{J.~J.}
  \bibnamefont{Bollinger}}, \bibinfo{journal}{Nature}
  \textbf{\bibinfo{volume}{484}}, \bibinfo{pages}{489} (\bibinfo{year}{2012}).

\bibitem[{\citenamefont{Sacha}(2015)}]{Sacha}
\bibinfo{author}{\bibfnamefont{K.}~\bibnamefont{Sacha}},
  \bibinfo{journal}{Phys. Rev. A} \textbf{\bibinfo{volume}{91}},
  \bibinfo{pages}{033617} (\bibinfo{year}{2015}),
  \urlprefix\url{https://link.aps.org/doi/10.1103/PhysRevA.91.033617}.

\bibitem[{\citenamefont{Else et~al.}(2016)\citenamefont{Else, Bauer, and
  Nayak}}]{ElseFTC}
\bibinfo{author}{\bibfnamefont{D.~V.} \bibnamefont{Else}},
  \bibinfo{author}{\bibfnamefont{B.}~\bibnamefont{Bauer}}, \bibnamefont{and}
  \bibinfo{author}{\bibfnamefont{C.}~\bibnamefont{Nayak}},
  \bibinfo{journal}{Phys. Rev. Lett.} \textbf{\bibinfo{volume}{117}},
  \bibinfo{pages}{090402} (\bibinfo{year}{2016}),
  \urlprefix\url{http://link.aps.org/doi/10.1103/PhysRevLett.117.090402}.

\bibitem[{\citenamefont{Khemani et~al.}(2016)\citenamefont{Khemani, Lazarides,
  Moessner, and Sondhi}}]{Khemani16}
\bibinfo{author}{\bibfnamefont{V.}~\bibnamefont{Khemani}},
  \bibinfo{author}{\bibfnamefont{A.}~\bibnamefont{Lazarides}},
  \bibinfo{author}{\bibfnamefont{R.}~\bibnamefont{Moessner}}, \bibnamefont{and}
  \bibinfo{author}{\bibfnamefont{S.~L.} \bibnamefont{Sondhi}},
  \bibinfo{journal}{Phys. Rev. Lett.} \textbf{\bibinfo{volume}{116}},
  \bibinfo{pages}{250401} (\bibinfo{year}{2016}),
  \urlprefix\url{http://link.aps.org/doi/10.1103/PhysRevLett.116.250401}.

\bibitem[{\citenamefont{Sacha and Zakrzewski}(2017)}]{sacha2017time}
\bibinfo{author}{\bibfnamefont{K.}~\bibnamefont{Sacha}} \bibnamefont{and}
  \bibinfo{author}{\bibfnamefont{J.}~\bibnamefont{Zakrzewski}},
  \bibinfo{journal}{Reports on Progress in Physics}
  \textbf{\bibinfo{volume}{81}}, \bibinfo{pages}{016401}
  (\bibinfo{year}{2017}).

\bibitem[{\citenamefont{Sacha}(2020)}]{sacha2020time}
\bibinfo{author}{\bibfnamefont{K.}~\bibnamefont{Sacha}},
  \emph{\bibinfo{title}{Time Crystals}} (\bibinfo{publisher}{Springer},
  \bibinfo{year}{2020}).

\bibitem[{\citenamefont{Zhang et~al.}(2017)\citenamefont{Zhang, Hess,
  Kyprianidis, Becker, Lee, Smith, Pagano, Potirniche, Potter, Vishwanath
  et~al.}}]{Monroe2017}
\bibinfo{author}{\bibfnamefont{J.}~\bibnamefont{Zhang}},
  \bibinfo{author}{\bibfnamefont{P.}~\bibnamefont{Hess}},
  \bibinfo{author}{\bibfnamefont{A.}~\bibnamefont{Kyprianidis}},
  \bibinfo{author}{\bibfnamefont{P.}~\bibnamefont{Becker}},
  \bibinfo{author}{\bibfnamefont{A.}~\bibnamefont{Lee}},
  \bibinfo{author}{\bibfnamefont{J.}~\bibnamefont{Smith}},
  \bibinfo{author}{\bibfnamefont{G.}~\bibnamefont{Pagano}},
  \bibinfo{author}{\bibfnamefont{I.-D.} \bibnamefont{Potirniche}},
  \bibinfo{author}{\bibfnamefont{A.~C.} \bibnamefont{Potter}},
  \bibinfo{author}{\bibfnamefont{A.}~\bibnamefont{Vishwanath}},
  \bibnamefont{et~al.}, \bibinfo{journal}{Nature}
  \textbf{\bibinfo{volume}{543}}, \bibinfo{pages}{217} (\bibinfo{year}{2017}).

\bibitem[{\citenamefont{Choi et~al.}(2017)\citenamefont{Choi, Choi, Landig,
  Kucsko, Zhou, Isoya, Jelezko, Onoda, Sumiya, Khemani
  et~al.}}]{choi2017observation}
\bibinfo{author}{\bibfnamefont{S.}~\bibnamefont{Choi}},
  \bibinfo{author}{\bibfnamefont{J.}~\bibnamefont{Choi}},
  \bibinfo{author}{\bibfnamefont{R.}~\bibnamefont{Landig}},
  \bibinfo{author}{\bibfnamefont{G.}~\bibnamefont{Kucsko}},
  \bibinfo{author}{\bibfnamefont{H.}~\bibnamefont{Zhou}},
  \bibinfo{author}{\bibfnamefont{J.}~\bibnamefont{Isoya}},
  \bibinfo{author}{\bibfnamefont{F.}~\bibnamefont{Jelezko}},
  \bibinfo{author}{\bibfnamefont{S.}~\bibnamefont{Onoda}},
  \bibinfo{author}{\bibfnamefont{H.}~\bibnamefont{Sumiya}},
  \bibinfo{author}{\bibfnamefont{V.}~\bibnamefont{Khemani}},
  \bibnamefont{et~al.}, \bibinfo{journal}{Nature}
  \textbf{\bibinfo{volume}{543}}, \bibinfo{pages}{221} (\bibinfo{year}{2017}).

\bibitem[{\citenamefont{Pal et~al.}(2018)\citenamefont{Pal, Nishad, Mahesh, and
  Sreejith}}]{Pal2018}
\bibinfo{author}{\bibfnamefont{S.}~\bibnamefont{Pal}},
  \bibinfo{author}{\bibfnamefont{N.}~\bibnamefont{Nishad}},
  \bibinfo{author}{\bibfnamefont{T.~S.} \bibnamefont{Mahesh}},
  \bibnamefont{and} \bibinfo{author}{\bibfnamefont{G.~J.}
  \bibnamefont{Sreejith}}, \bibinfo{journal}{Phys. Rev. Lett.}
  \textbf{\bibinfo{volume}{120}}, \bibinfo{pages}{180602}
  (\bibinfo{year}{2018}),
  \urlprefix\url{https://link.aps.org/doi/10.1103/PhysRevLett.120.180602}.

\bibitem[{\citenamefont{Rovny et~al.}(2018)\citenamefont{Rovny, Blum, and
  Barrett}}]{Rovny2018}
\bibinfo{author}{\bibfnamefont{J.}~\bibnamefont{Rovny}},
  \bibinfo{author}{\bibfnamefont{R.~L.} \bibnamefont{Blum}}, \bibnamefont{and}
  \bibinfo{author}{\bibfnamefont{S.~E.} \bibnamefont{Barrett}},
  \bibinfo{journal}{Phys. Rev. Lett.} \textbf{\bibinfo{volume}{120}},
  \bibinfo{pages}{180603} (\bibinfo{year}{2018}),
  \urlprefix\url{https://link.aps.org/doi/10.1103/PhysRevLett.120.180603}.

\bibitem[{\citenamefont{Smits et~al.}(2018)\citenamefont{Smits, Liao, Stoof,
  and van~der Straten}}]{smits2018observation}
\bibinfo{author}{\bibfnamefont{J.}~\bibnamefont{Smits}},
  \bibinfo{author}{\bibfnamefont{L.}~\bibnamefont{Liao}},
  \bibinfo{author}{\bibfnamefont{H.}~\bibnamefont{Stoof}}, \bibnamefont{and}
  \bibinfo{author}{\bibfnamefont{P.}~\bibnamefont{van~der Straten}},
  \bibinfo{journal}{Physical review letters} \textbf{\bibinfo{volume}{121}},
  \bibinfo{pages}{185301} (\bibinfo{year}{2018}).

\bibitem[{\citenamefont{Giergiel
  et~al.}(2018{\natexlab{a}})\citenamefont{Giergiel, Kosior, Hannaford, and
  Sacha}}]{giergiel2018time}
\bibinfo{author}{\bibfnamefont{K.}~\bibnamefont{Giergiel}},
  \bibinfo{author}{\bibfnamefont{A.}~\bibnamefont{Kosior}},
  \bibinfo{author}{\bibfnamefont{P.}~\bibnamefont{Hannaford}},
  \bibnamefont{and} \bibinfo{author}{\bibfnamefont{K.}~\bibnamefont{Sacha}},
  \bibinfo{journal}{Physical Review A} \textbf{\bibinfo{volume}{98}},
  \bibinfo{pages}{013613} (\bibinfo{year}{2018}{\natexlab{a}}).

\bibitem[{\citenamefont{Surace et~al.}(2019)\citenamefont{Surace, Russomanno,
  Dalmonte, Silva, Fazio, and Iemini}}]{surace2019floquet}
\bibinfo{author}{\bibfnamefont{F.~M.} \bibnamefont{Surace}},
  \bibinfo{author}{\bibfnamefont{A.}~\bibnamefont{Russomanno}},
  \bibinfo{author}{\bibfnamefont{M.}~\bibnamefont{Dalmonte}},
  \bibinfo{author}{\bibfnamefont{A.}~\bibnamefont{Silva}},
  \bibinfo{author}{\bibfnamefont{R.}~\bibnamefont{Fazio}}, \bibnamefont{and}
  \bibinfo{author}{\bibfnamefont{F.}~\bibnamefont{Iemini}},
  \bibinfo{journal}{Physical Review B} \textbf{\bibinfo{volume}{99}},
  \bibinfo{pages}{104303} (\bibinfo{year}{2019}).

\bibitem[{\citenamefont{Pizzi et~al.}(2019)\citenamefont{Pizzi, Knolle, and
  Nunnenkamp}}]{pizzi2019period}
\bibinfo{author}{\bibfnamefont{A.}~\bibnamefont{Pizzi}},
  \bibinfo{author}{\bibfnamefont{J.}~\bibnamefont{Knolle}}, \bibnamefont{and}
  \bibinfo{author}{\bibfnamefont{A.}~\bibnamefont{Nunnenkamp}},
  \bibinfo{journal}{Physical review letters} \textbf{\bibinfo{volume}{123}},
  \bibinfo{pages}{150601} (\bibinfo{year}{2019}).

\bibitem[{\citenamefont{Kuro{\'{s}} et~al.}(2020)\citenamefont{Kuro{\'{s}},
  Mukherjee, Golletz, Sauvage, Giergiel, Mintert, and Sacha}}]{Kuro__2020}
\bibinfo{author}{\bibfnamefont{A.}~\bibnamefont{Kuro{\'{s}}}},
  \bibinfo{author}{\bibfnamefont{R.}~\bibnamefont{Mukherjee}},
  \bibinfo{author}{\bibfnamefont{W.}~\bibnamefont{Golletz}},
  \bibinfo{author}{\bibfnamefont{F.}~\bibnamefont{Sauvage}},
  \bibinfo{author}{\bibfnamefont{K.}~\bibnamefont{Giergiel}},
  \bibinfo{author}{\bibfnamefont{F.}~\bibnamefont{Mintert}}, \bibnamefont{and}
  \bibinfo{author}{\bibfnamefont{K.}~\bibnamefont{Sacha}},
  \bibinfo{journal}{New Journal of Physics} \textbf{\bibinfo{volume}{22}},
  \bibinfo{pages}{095001} (\bibinfo{year}{2020}),
  \urlprefix\url{https://doi.org/10.1088/1367-2630/abb03e}.

\bibitem[{\citenamefont{Giergiel et~al.}(2020)\citenamefont{Giergiel, Tran,
  Zaheer, Singh, Sidorov, Sacha, and Hannaford}}]{giergiel2020creating}
\bibinfo{author}{\bibfnamefont{K.}~\bibnamefont{Giergiel}},
  \bibinfo{author}{\bibfnamefont{T.}~\bibnamefont{Tran}},
  \bibinfo{author}{\bibfnamefont{A.}~\bibnamefont{Zaheer}},
  \bibinfo{author}{\bibfnamefont{A.}~\bibnamefont{Singh}},
  \bibinfo{author}{\bibfnamefont{A.}~\bibnamefont{Sidorov}},
  \bibinfo{author}{\bibfnamefont{K.}~\bibnamefont{Sacha}}, \bibnamefont{and}
  \bibinfo{author}{\bibfnamefont{P.}~\bibnamefont{Hannaford}},
  \bibinfo{journal}{New Journal of Physics} \textbf{\bibinfo{volume}{22}},
  \bibinfo{pages}{085004} (\bibinfo{year}{2020}).

\bibitem[{\citenamefont{Pizzi et~al.}(2021)\citenamefont{Pizzi, Knolle, and
  Nunnenkamp}}]{Pizzi2021}
\bibinfo{author}{\bibfnamefont{A.}~\bibnamefont{Pizzi}},
  \bibinfo{author}{\bibfnamefont{J.}~\bibnamefont{Knolle}}, \bibnamefont{and}
  \bibinfo{author}{\bibfnamefont{A.}~\bibnamefont{Nunnenkamp}},
  \bibinfo{journal}{Nature Communications} \textbf{\bibinfo{volume}{12}},
  \bibinfo{pages}{2341} (\bibinfo{year}{2021}), ISSN \bibinfo{issn}{2041-1723},
  \urlprefix\url{https://doi.org/10.1038/s41467-021-22583-5}.

\bibitem[{\citenamefont{Guo et~al.}(2013)\citenamefont{Guo, Marthaler, and
  Sch{\"o}n}}]{guo2013phase}
\bibinfo{author}{\bibfnamefont{L.}~\bibnamefont{Guo}},
  \bibinfo{author}{\bibfnamefont{M.}~\bibnamefont{Marthaler}},
  \bibnamefont{and}
  \bibinfo{author}{\bibfnamefont{G.}~\bibnamefont{Sch{\"o}n}},
  \bibinfo{journal}{Physical review letters} \textbf{\bibinfo{volume}{111}},
  \bibinfo{pages}{205303} (\bibinfo{year}{2013}).

\bibitem[{\citenamefont{Giergiel
  et~al.}(2018{\natexlab{b}})\citenamefont{Giergiel, Miroszewski, and
  Sacha}}]{Giergiel}
\bibinfo{author}{\bibfnamefont{K.}~\bibnamefont{Giergiel}},
  \bibinfo{author}{\bibfnamefont{A.}~\bibnamefont{Miroszewski}},
  \bibnamefont{and} \bibinfo{author}{\bibfnamefont{K.}~\bibnamefont{Sacha}},
  \bibinfo{journal}{Phys. Rev. Lett.} \textbf{\bibinfo{volume}{120}},
  \bibinfo{pages}{140401} (\bibinfo{year}{2018}{\natexlab{b}}),
  \urlprefix\url{https://link.aps.org/doi/10.1103/PhysRevLett.120.140401}.

\bibitem[{\citenamefont{Pengfei et~al.}(2018)\citenamefont{Pengfei, Michael,
  and Guo}}]{Liang2017}
\bibinfo{author}{\bibfnamefont{L.}~\bibnamefont{Pengfei}},
  \bibinfo{author}{\bibfnamefont{M.}~\bibnamefont{Michael}}, \bibnamefont{and}
  \bibinfo{author}{\bibfnamefont{L.}~\bibnamefont{Guo}}, \bibinfo{journal}{New
  Journal of Physics} \textbf{\bibinfo{volume}{20}}, \bibinfo{pages}{023043}
  (\bibinfo{year}{2018}), ISSN \bibinfo{issn}{1367-2630},
  \urlprefix\url{http://stacks.iop.org/1367-2630/20/i=2/a=023043}.

\bibitem[{\citenamefont{Giergiel
  et~al.}(2019{\natexlab{a}})\citenamefont{Giergiel, Kuro\'s, and
  Sacha}}]{Giergiel2019}
\bibinfo{author}{\bibfnamefont{K.}~\bibnamefont{Giergiel}},
  \bibinfo{author}{\bibfnamefont{A.}~\bibnamefont{Kuro\'s}}, \bibnamefont{and}
  \bibinfo{author}{\bibfnamefont{K.}~\bibnamefont{Sacha}},
  \bibinfo{journal}{Phys. Rev. B} \textbf{\bibinfo{volume}{99}},
  \bibinfo{pages}{220303} (\bibinfo{year}{2019}{\natexlab{a}}),
  \urlprefix\url{https://link.aps.org/doi/10.1103/PhysRevB.99.220303}.

\bibitem[{\citenamefont{Guo and Liang}(2020)}]{guo2020condensed}
\bibinfo{author}{\bibfnamefont{L.}~\bibnamefont{Guo}} \bibnamefont{and}
  \bibinfo{author}{\bibfnamefont{P.}~\bibnamefont{Liang}},
  \bibinfo{journal}{New Journal of Physics} \textbf{\bibinfo{volume}{22}},
  \bibinfo{pages}{075003} (\bibinfo{year}{2020}).

\bibitem[{\citenamefont{Zlabys et~al.}(2021)\citenamefont{Zlabys, Fan,
  Anisimovas, and Sacha}}]{Giedrius}
\bibinfo{author}{\bibfnamefont{G.}~\bibnamefont{Zlabys}},
  \bibinfo{author}{\bibfnamefont{C.-h.} \bibnamefont{Fan}},
  \bibinfo{author}{\bibfnamefont{E.}~\bibnamefont{Anisimovas}},
  \bibnamefont{and} \bibinfo{author}{\bibfnamefont{K.}~\bibnamefont{Sacha}},
  \bibinfo{journal}{Physical Review B} \textbf{\bibinfo{volume}{103}},
  \bibinfo{pages}{L100301} (\bibinfo{year}{2021}).

\bibitem[{\citenamefont{Taheri et~al.}(2020)\citenamefont{Taheri, Matsko,
  Maleki, and Sacha}}]{taheri2020all}
\bibinfo{author}{\bibfnamefont{H.}~\bibnamefont{Taheri}},
  \bibinfo{author}{\bibfnamefont{A.~B.} \bibnamefont{Matsko}},
  \bibinfo{author}{\bibfnamefont{L.}~\bibnamefont{Maleki}}, \bibnamefont{and}
  \bibinfo{author}{\bibfnamefont{K.}~\bibnamefont{Sacha}},
  \bibinfo{journal}{arXiv preprint arXiv:2012.07927}  (\bibinfo{year}{2020}).

\bibitem[{\citenamefont{Yang and Cai}(2021)}]{yang2021dynamical}
\bibinfo{author}{\bibfnamefont{X.}~\bibnamefont{Yang}} \bibnamefont{and}
  \bibinfo{author}{\bibfnamefont{Z.}~\bibnamefont{Cai}},
  \bibinfo{journal}{Physical Review Letters} \textbf{\bibinfo{volume}{126}},
  \bibinfo{pages}{020602} (\bibinfo{year}{2021}).

\bibitem[{\citenamefont{Sakurai et~al.}(2021)\citenamefont{Sakurai, Bastidas,
  Munro, and Nemoto}}]{sakurai2021chimera}
\bibinfo{author}{\bibfnamefont{A.}~\bibnamefont{Sakurai}},
  \bibinfo{author}{\bibfnamefont{V.}~\bibnamefont{Bastidas}},
  \bibinfo{author}{\bibfnamefont{W.}~\bibnamefont{Munro}}, \bibnamefont{and}
  \bibinfo{author}{\bibfnamefont{K.}~\bibnamefont{Nemoto}},
  \bibinfo{journal}{Physical Review Letters} \textbf{\bibinfo{volume}{126}},
  \bibinfo{pages}{120606} (\bibinfo{year}{2021}).

\bibitem[{\citenamefont{Buchleitner et~al.}(2002)\citenamefont{Buchleitner,
  Delande, and Zakrzewski}}]{buchleitner2002non}
\bibinfo{author}{\bibfnamefont{A.}~\bibnamefont{Buchleitner}},
  \bibinfo{author}{\bibfnamefont{D.}~\bibnamefont{Delande}}, \bibnamefont{and}
  \bibinfo{author}{\bibfnamefont{J.}~\bibnamefont{Zakrzewski}},
  \bibinfo{journal}{Physics reports} \textbf{\bibinfo{volume}{368}},
  \bibinfo{pages}{409} (\bibinfo{year}{2002}).

\bibitem[{\citenamefont{{Wang} et~al.}(2021)\citenamefont{{Wang}, {Sacha},
  {Hannaford}, and {Dalton}}}]{Wang2021two-mode}
\bibinfo{author}{\bibfnamefont{J.}~\bibnamefont{{Wang}}},
  \bibinfo{author}{\bibfnamefont{K.}~\bibnamefont{{Sacha}}},
  \bibinfo{author}{\bibfnamefont{P.}~\bibnamefont{{Hannaford}}},
  \bibnamefont{and} \bibinfo{author}{\bibfnamefont{B.~J.}
  \bibnamefont{{Dalton}}}, \bibinfo{journal}{arXiv e-prints}
  \bibinfo{eid}{arXiv:2106.02219} (\bibinfo{year}{2021}), \eprint{2106.02219}.

\bibitem[{\citenamefont{Pethick and Smith}(2002)}]{Pethick2002}
\bibinfo{author}{\bibfnamefont{C.}~\bibnamefont{Pethick}} \bibnamefont{and}
  \bibinfo{author}{\bibfnamefont{H.}~\bibnamefont{Smith}},
  \emph{\bibinfo{title}{{Bose-Eistein condensation in dilute gases}}}
  (\bibinfo{publisher}{{Cambridge University Press}},
  \bibinfo{address}{{Cambridge, England}}, \bibinfo{year}{2002}).

\bibitem[{\citenamefont{Mukherjee
  et~al.}(2020{\natexlab{a}})\citenamefont{Mukherjee, Xie, and
  Mintert}}]{Mukherjee2}
\bibinfo{author}{\bibfnamefont{R.}~\bibnamefont{Mukherjee}},
  \bibinfo{author}{\bibfnamefont{H.}~\bibnamefont{Xie}}, \bibnamefont{and}
  \bibinfo{author}{\bibfnamefont{F.}~\bibnamefont{Mintert}},
  \bibinfo{journal}{Phys. Rev. Lett.} \textbf{\bibinfo{volume}{125}},
  \bibinfo{pages}{203603} (\bibinfo{year}{2020}{\natexlab{a}}),
  \urlprefix\url{https://link.aps.org/doi/10.1103/PhysRevLett.125.203603}.

\bibitem[{\citenamefont{Sauvage and Mintert}(2020)}]{Sauvage}
\bibinfo{author}{\bibfnamefont{F.}~\bibnamefont{Sauvage}} \bibnamefont{and}
  \bibinfo{author}{\bibfnamefont{F.}~\bibnamefont{Mintert}},
  \bibinfo{journal}{PRX Quantum} \textbf{\bibinfo{volume}{1}},
  \bibinfo{pages}{020322} (\bibinfo{year}{2020}),
  \urlprefix\url{https://link.aps.org/doi/10.1103/PRXQuantum.1.020322}.

\bibitem[{gpy(2016)}]{gpyopt2016}
\emph{\bibinfo{title}{Gpyopt: A bayesian optimization framework in python}}
  (\bibinfo{year}{2016}), \urlprefix\url{http://github.com/SheffieldML/GPyOpt}.

\bibitem[{\citenamefont{Snoek et~al.}(2012)\citenamefont{Snoek, Larochelle, and
  Adams}}]{snoek2012practical}
\bibinfo{author}{\bibfnamefont{J.}~\bibnamefont{Snoek}},
  \bibinfo{author}{\bibfnamefont{H.}~\bibnamefont{Larochelle}},
  \bibnamefont{and} \bibinfo{author}{\bibfnamefont{R.~P.} \bibnamefont{Adams}},
  pp. \bibinfo{pages}{2951--2959} (\bibinfo{year}{2012}).

\bibitem[{\citenamefont{Frazier}(2018)}]{frazier2018tutorial}
\bibinfo{author}{\bibfnamefont{P.~I.} \bibnamefont{Frazier}},
  \bibinfo{journal}{arXiv:1807.02811}  (\bibinfo{year}{2018}),
  \urlprefix\url{https://arxiv.org/abs/1807.02811}.

\bibitem[{\citenamefont{Wang et~al.}(2021)\citenamefont{Wang, Hannaford, and
  Dalton}}]{wang2021many}
\bibinfo{author}{\bibfnamefont{J.}~\bibnamefont{Wang}},
  \bibinfo{author}{\bibfnamefont{P.}~\bibnamefont{Hannaford}},
  \bibnamefont{and} \bibinfo{author}{\bibfnamefont{B.~J.}
  \bibnamefont{Dalton}}, \bibinfo{journal}{New Journal of Physics}
  (\bibinfo{year}{2021}).

\bibitem[{\citenamefont{Steane et~al.}(1995)\citenamefont{Steane, Szriftgiser,
  Desbiolles, and Dalibard}}]{Steane95}
\bibinfo{author}{\bibfnamefont{A.}~\bibnamefont{Steane}},
  \bibinfo{author}{\bibfnamefont{P.}~\bibnamefont{Szriftgiser}},
  \bibinfo{author}{\bibfnamefont{P.}~\bibnamefont{Desbiolles}},
  \bibnamefont{and} \bibinfo{author}{\bibfnamefont{J.}~\bibnamefont{Dalibard}},
  \bibinfo{journal}{Phys. Rev. Lett.} \textbf{\bibinfo{volume}{74}},
  \bibinfo{pages}{4972} (\bibinfo{year}{1995}),
  \urlprefix\url{http://link.aps.org/doi/10.1103/PhysRevLett.74.4972}.

\bibitem[{\citenamefont{Roach et~al.}(1995)\citenamefont{Roach, Abele, Boshier,
  Grossman, Zetie, and Hinds}}]{Roach1995}
\bibinfo{author}{\bibfnamefont{T.~M.} \bibnamefont{Roach}},
  \bibinfo{author}{\bibfnamefont{H.}~\bibnamefont{Abele}},
  \bibinfo{author}{\bibfnamefont{M.~G.} \bibnamefont{Boshier}},
  \bibinfo{author}{\bibfnamefont{H.~L.} \bibnamefont{Grossman}},
  \bibinfo{author}{\bibfnamefont{K.~P.} \bibnamefont{Zetie}}, \bibnamefont{and}
  \bibinfo{author}{\bibfnamefont{E.~A.} \bibnamefont{Hinds}},
  \bibinfo{journal}{Phys. Rev. Lett.} \textbf{\bibinfo{volume}{75}},
  \bibinfo{pages}{629} (\bibinfo{year}{1995}),
  \urlprefix\url{https://link.aps.org/doi/10.1103/PhysRevLett.75.629}.

\bibitem[{\citenamefont{Sidorov et~al.}(1996)\citenamefont{Sidorov, McLean,
  Rowlands, Lau, Murphy, Walkiewicz, Opat, and Hannaford}}]{Sidorov1996}
\bibinfo{author}{\bibfnamefont{A.~I.} \bibnamefont{Sidorov}},
  \bibinfo{author}{\bibfnamefont{R.~J.} \bibnamefont{McLean}},
  \bibinfo{author}{\bibfnamefont{W.~J.} \bibnamefont{Rowlands}},
  \bibinfo{author}{\bibfnamefont{D.~C.} \bibnamefont{Lau}},
  \bibinfo{author}{\bibfnamefont{J.~E.} \bibnamefont{Murphy}},
  \bibinfo{author}{\bibfnamefont{M.}~\bibnamefont{Walkiewicz}},
  \bibinfo{author}{\bibfnamefont{G.~I.} \bibnamefont{Opat}}, \bibnamefont{and}
  \bibinfo{author}{\bibfnamefont{P.}~\bibnamefont{Hannaford}},
  \bibinfo{journal}{Quantum and Semiclassical Optics: Journal of the European
  Optical Society Part B} \textbf{\bibinfo{volume}{8}}, \bibinfo{pages}{713}
  (\bibinfo{year}{1996}),
  \urlprefix\url{http://stacks.iop.org/1355-5111/8/i=3/a=030}.

\bibitem[{\citenamefont{Westbrook et~al.}(1998)\citenamefont{Westbrook,
  Westbrook, Landragin, Labeyrie, Cognet, Savalli, Horvath, Aspect, Hendel,
  Moelmer et~al.}}]{Westbrook1998}
\bibinfo{author}{\bibfnamefont{N.}~\bibnamefont{Westbrook}},
  \bibinfo{author}{\bibfnamefont{C.~I.} \bibnamefont{Westbrook}},
  \bibinfo{author}{\bibfnamefont{A.}~\bibnamefont{Landragin}},
  \bibinfo{author}{\bibfnamefont{G.}~\bibnamefont{Labeyrie}},
  \bibinfo{author}{\bibfnamefont{L.}~\bibnamefont{Cognet}},
  \bibinfo{author}{\bibfnamefont{V.}~\bibnamefont{Savalli}},
  \bibinfo{author}{\bibfnamefont{G.}~\bibnamefont{Horvath}},
  \bibinfo{author}{\bibfnamefont{A.}~\bibnamefont{Aspect}},
  \bibinfo{author}{\bibfnamefont{C.}~\bibnamefont{Hendel}},
  \bibinfo{author}{\bibfnamefont{K.}~\bibnamefont{Moelmer}},
  \bibnamefont{et~al.}, \bibinfo{journal}{Physica Scripta}
  \textbf{\bibinfo{volume}{1998}}, \bibinfo{pages}{7} (\bibinfo{year}{1998}),
  \urlprefix\url{http://stacks.iop.org/1402-4896/1998/i=T78/a=001}.

\bibitem[{\citenamefont{Lau et~al.}(1999)\citenamefont{Lau, Sidorov, Opat,
  McLean, Rowlands, and Hannaford}}]{Lau1999}
\bibinfo{author}{\bibfnamefont{D.~C.} \bibnamefont{Lau}},
  \bibinfo{author}{\bibfnamefont{A.~I.} \bibnamefont{Sidorov}},
  \bibinfo{author}{\bibfnamefont{G.~I.} \bibnamefont{Opat}},
  \bibinfo{author}{\bibfnamefont{R.~J.} \bibnamefont{McLean}},
  \bibinfo{author}{\bibfnamefont{W.~J.} \bibnamefont{Rowlands}},
  \bibnamefont{and}
  \bibinfo{author}{\bibfnamefont{P.}~\bibnamefont{Hannaford}},
  \bibinfo{journal}{Eur. Phys. J. D} \textbf{\bibinfo{volume}{5}},
  \bibinfo{pages}{193} (\bibinfo{year}{1999}),
  \urlprefix\url{https://doi.org/10.1007/s100530050244}.

\bibitem[{\citenamefont{Bongs et~al.}(1999{\natexlab{a}})\citenamefont{Bongs,
  Burger, Birkl, Sengstock, Ertmer, Rz\c{a}\.zewski, Sanpera, and
  Lewenstein}}]{Bongs1999}
\bibinfo{author}{\bibfnamefont{K.}~\bibnamefont{Bongs}},
  \bibinfo{author}{\bibfnamefont{S.}~\bibnamefont{Burger}},
  \bibinfo{author}{\bibfnamefont{G.}~\bibnamefont{Birkl}},
  \bibinfo{author}{\bibfnamefont{K.}~\bibnamefont{Sengstock}},
  \bibinfo{author}{\bibfnamefont{W.}~\bibnamefont{Ertmer}},
  \bibinfo{author}{\bibfnamefont{K.}~\bibnamefont{Rz\c{a}\.zewski}},
  \bibinfo{author}{\bibfnamefont{A.}~\bibnamefont{Sanpera}}, \bibnamefont{and}
  \bibinfo{author}{\bibfnamefont{M.}~\bibnamefont{Lewenstein}},
  \bibinfo{journal}{Phys. Rev. Lett.} \textbf{\bibinfo{volume}{83}},
  \bibinfo{pages}{3577} (\bibinfo{year}{1999}{\natexlab{a}}),
  \urlprefix\url{https://link.aps.org/doi/10.1103/PhysRevLett.83.3577}.

\bibitem[{\citenamefont{Sidorov et~al.}(2002)\citenamefont{Sidorov, McLean,
  Scharnberg, Gough, Davis, Sexton, Opat, and Hannaford}}]{Sidorov2002}
\bibinfo{author}{\bibfnamefont{A.}~\bibnamefont{Sidorov}},
  \bibinfo{author}{\bibfnamefont{R.}~\bibnamefont{McLean}},
  \bibinfo{author}{\bibfnamefont{F.}~\bibnamefont{Scharnberg}},
  \bibinfo{author}{\bibfnamefont{D.}~\bibnamefont{Gough}},
  \bibinfo{author}{\bibfnamefont{T.}~\bibnamefont{Davis}},
  \bibinfo{author}{\bibfnamefont{B.}~\bibnamefont{Sexton}},
  \bibinfo{author}{\bibfnamefont{G.}~\bibnamefont{Opat}}, \bibnamefont{and}
  \bibinfo{author}{\bibfnamefont{P.}~\bibnamefont{Hannaford}},
  \bibinfo{journal}{Acta Phys. Pol. B} \textbf{\bibinfo{volume}{33}},
  \bibinfo{pages}{2137} (\bibinfo{year}{2002}).

\bibitem[{\citenamefont{Fiutowski et~al.}(2013)\citenamefont{Fiutowski,
  Bartoszek-Bober, Dohnalik, and Kawalec}}]{Fiutowski2013}
\bibinfo{author}{\bibfnamefont{J.}~\bibnamefont{Fiutowski}},
  \bibinfo{author}{\bibfnamefont{D.}~\bibnamefont{Bartoszek-Bober}},
  \bibinfo{author}{\bibfnamefont{T.}~\bibnamefont{Dohnalik}}, \bibnamefont{and}
  \bibinfo{author}{\bibfnamefont{T.}~\bibnamefont{Kawalec}},
  \bibinfo{journal}{Optics Communications} \textbf{\bibinfo{volume}{297}},
  \bibinfo{pages}{59 } (\bibinfo{year}{2013}), ISSN \bibinfo{issn}{0030-4018},
  \urlprefix\url{http://www.sciencedirect.com/science/article/pii/S0030401813001521}.

\bibitem[{\citenamefont{Kawalec et~al.}(2014)\citenamefont{Kawalec,
  Bartoszek-Bober, Pana\'{s}, Fiutowski, P{\l}awecka, and
  Rubahn}}]{Kawalec2014}
\bibinfo{author}{\bibfnamefont{T.}~\bibnamefont{Kawalec}},
  \bibinfo{author}{\bibfnamefont{D.}~\bibnamefont{Bartoszek-Bober}},
  \bibinfo{author}{\bibfnamefont{R.}~\bibnamefont{Pana\'{s}}},
  \bibinfo{author}{\bibfnamefont{J.}~\bibnamefont{Fiutowski}},
  \bibinfo{author}{\bibfnamefont{A.}~\bibnamefont{P{\l}awecka}},
  \bibnamefont{and} \bibinfo{author}{\bibfnamefont{H.-G.}
  \bibnamefont{Rubahn}}, \bibinfo{journal}{Opt. Lett.}
  \textbf{\bibinfo{volume}{39}}, \bibinfo{pages}{2932} (\bibinfo{year}{2014}),
  \urlprefix\url{http://ol.osa.org/abstract.cfm?URI=ol-39-10-2932}.

\bibitem[{\citenamefont{Hadzibabic et~al.}(2006)\citenamefont{Hadzibabic,
  Kr{\"u}ger, Cheneau, Battelier, and Dalibard}}]{Hadzibabic}
\bibinfo{author}{\bibfnamefont{Z.}~\bibnamefont{Hadzibabic}},
  \bibinfo{author}{\bibfnamefont{P.}~\bibnamefont{Kr{\"u}ger}},
  \bibinfo{author}{\bibfnamefont{M.}~\bibnamefont{Cheneau}},
  \bibinfo{author}{\bibfnamefont{B.}~\bibnamefont{Battelier}},
  \bibnamefont{and} \bibinfo{author}{\bibfnamefont{J.}~\bibnamefont{Dalibard}},
  \bibinfo{journal}{Nature} \textbf{\bibinfo{volume}{441}},
  \bibinfo{pages}{1118} (\bibinfo{year}{2006}).

\bibitem[{\citenamefont{Amo et~al.}(2009)\citenamefont{Amo, Lefr{\`e}re,
  Pigeon, Adrados, Ciuti, Carusotto, Houdr{\'e}, Giacobino, and Bramati}}]{Amo}
\bibinfo{author}{\bibfnamefont{A.}~\bibnamefont{Amo}},
  \bibinfo{author}{\bibfnamefont{J.}~\bibnamefont{Lefr{\`e}re}},
  \bibinfo{author}{\bibfnamefont{S.}~\bibnamefont{Pigeon}},
  \bibinfo{author}{\bibfnamefont{C.}~\bibnamefont{Adrados}},
  \bibinfo{author}{\bibfnamefont{C.}~\bibnamefont{Ciuti}},
  \bibinfo{author}{\bibfnamefont{I.}~\bibnamefont{Carusotto}},
  \bibinfo{author}{\bibfnamefont{R.}~\bibnamefont{Houdr{\'e}}},
  \bibinfo{author}{\bibfnamefont{E.}~\bibnamefont{Giacobino}},
  \bibnamefont{and} \bibinfo{author}{\bibfnamefont{A.}~\bibnamefont{Bramati}},
  \bibinfo{journal}{Nature Physics} \textbf{\bibinfo{volume}{5}},
  \bibinfo{pages}{805} (\bibinfo{year}{2009}).

\bibitem[{\citenamefont{Weiler et~al.}(2008)\citenamefont{Weiler, Neely,
  Scherer, Bradley, Davis, and Anderson}}]{Weiler}
\bibinfo{author}{\bibfnamefont{C.~N.} \bibnamefont{Weiler}},
  \bibinfo{author}{\bibfnamefont{T.~W.} \bibnamefont{Neely}},
  \bibinfo{author}{\bibfnamefont{D.~R.} \bibnamefont{Scherer}},
  \bibinfo{author}{\bibfnamefont{A.~S.} \bibnamefont{Bradley}},
  \bibinfo{author}{\bibfnamefont{M.~J.} \bibnamefont{Davis}}, \bibnamefont{and}
  \bibinfo{author}{\bibfnamefont{B.~P.} \bibnamefont{Anderson}},
  \bibinfo{journal}{Nature} \textbf{\bibinfo{volume}{455}},
  \bibinfo{pages}{948} (\bibinfo{year}{2008}).

\bibitem[{\citenamefont{Lustig et~al.}(2018)\citenamefont{Lustig, Sharabi, and
  Segev}}]{lustig2018topological}
\bibinfo{author}{\bibfnamefont{E.}~\bibnamefont{Lustig}},
  \bibinfo{author}{\bibfnamefont{Y.}~\bibnamefont{Sharabi}}, \bibnamefont{and}
  \bibinfo{author}{\bibfnamefont{M.}~\bibnamefont{Segev}},
  \bibinfo{journal}{Optica} \textbf{\bibinfo{volume}{5}}, \bibinfo{pages}{1390}
  (\bibinfo{year}{2018}).

\bibitem[{\citenamefont{Liang et~al.}(2018)\citenamefont{Liang, Marthaler, and
  Guo}}]{liang2018floquet}
\bibinfo{author}{\bibfnamefont{P.}~\bibnamefont{Liang}},
  \bibinfo{author}{\bibfnamefont{M.}~\bibnamefont{Marthaler}},
  \bibnamefont{and} \bibinfo{author}{\bibfnamefont{L.}~\bibnamefont{Guo}},
  \bibinfo{journal}{New Journal of Physics} \textbf{\bibinfo{volume}{20}},
  \bibinfo{pages}{023043} (\bibinfo{year}{2018}).

\bibitem[{\citenamefont{Giergiel
  et~al.}(2019{\natexlab{b}})\citenamefont{Giergiel, Dauphin, Lewenstein,
  Zakrzewski, and Sacha}}]{Giergiel_2019}
\bibinfo{author}{\bibfnamefont{K.}~\bibnamefont{Giergiel}},
  \bibinfo{author}{\bibfnamefont{A.}~\bibnamefont{Dauphin}},
  \bibinfo{author}{\bibfnamefont{M.}~\bibnamefont{Lewenstein}},
  \bibinfo{author}{\bibfnamefont{J.}~\bibnamefont{Zakrzewski}},
  \bibnamefont{and} \bibinfo{author}{\bibfnamefont{K.}~\bibnamefont{Sacha}},
  \bibinfo{journal}{New Journal of Physics} \textbf{\bibinfo{volume}{21}},
  \bibinfo{pages}{052003} (\bibinfo{year}{2019}{\natexlab{b}}),
  \urlprefix\url{https://doi.org/10.1088/1367-2630/ab1e5f}.

\bibitem[{\citenamefont{Giergiel et~al.}(2021)\citenamefont{Giergiel, Kuroś,
  Kosior, and Sacha}}]{giergiel2021inseparable}
\bibinfo{author}{\bibfnamefont{K.}~\bibnamefont{Giergiel}},
  \bibinfo{author}{\bibfnamefont{A.}~\bibnamefont{Kuroś}},
  \bibinfo{author}{\bibfnamefont{A.}~\bibnamefont{Kosior}}, \bibnamefont{and}
  \bibinfo{author}{\bibfnamefont{K.}~\bibnamefont{Sacha}},
  \emph{\bibinfo{title}{Inseparable time-crystal geometries on the m\"obius
  strip}} (\bibinfo{year}{2021}), \eprint{2103.03778}.

\bibitem[{\citenamefont{Mukherjee
  et~al.}(2020{\natexlab{b}})\citenamefont{Mukherjee, Sauvage, Xie, Löw, and
  Mintert}}]{Mukherjee_2020}
\bibinfo{author}{\bibfnamefont{R.}~\bibnamefont{Mukherjee}},
  \bibinfo{author}{\bibfnamefont{F.}~\bibnamefont{Sauvage}},
  \bibinfo{author}{\bibfnamefont{H.}~\bibnamefont{Xie}},
  \bibinfo{author}{\bibfnamefont{R.}~\bibnamefont{Löw}}, \bibnamefont{and}
  \bibinfo{author}{\bibfnamefont{F.}~\bibnamefont{Mintert}},
  \bibinfo{journal}{New Journal of Physics} \textbf{\bibinfo{volume}{22}},
  \bibinfo{pages}{075001} (\bibinfo{year}{2020}{\natexlab{b}}),
  \urlprefix\url{https://doi.org/10.1088%2F1367-2630%2Fab8677}.

\bibitem[{\citenamefont{Torrontegui et~al.}(2013)\citenamefont{Torrontegui,
  Ibáñez, Martínez-Garaot, Modugno, {del Campo}, Guéry-Odelin, Ruschhaupt,
  Chen, and Muga}}]{TORRONTEGUI2013117}
\bibinfo{author}{\bibfnamefont{E.}~\bibnamefont{Torrontegui}},
  \bibinfo{author}{\bibfnamefont{S.}~\bibnamefont{Ibáñez}},
  \bibinfo{author}{\bibfnamefont{S.}~\bibnamefont{Martínez-Garaot}},
  \bibinfo{author}{\bibfnamefont{M.}~\bibnamefont{Modugno}},
  \bibinfo{author}{\bibfnamefont{A.}~\bibnamefont{{del Campo}}},
  \bibinfo{author}{\bibfnamefont{D.}~\bibnamefont{Guéry-Odelin}},
  \bibinfo{author}{\bibfnamefont{A.}~\bibnamefont{Ruschhaupt}},
  \bibinfo{author}{\bibfnamefont{X.}~\bibnamefont{Chen}}, \bibnamefont{and}
  \bibinfo{author}{\bibfnamefont{J.~G.} \bibnamefont{Muga}},
  \textbf{\bibinfo{volume}{62}}, \bibinfo{pages}{117} (\bibinfo{year}{2013}),
  ISSN \bibinfo{issn}{1049-250X},
  \urlprefix\url{https://www.sciencedirect.com/science/article/pii/B9780124080904000025}.

\bibitem[{\citenamefont{{Zhou} et~al.}(2018)\citenamefont{{Zhou}, {Jin}, and
  {Schmiedmayer}}}]{Zhou}
\bibinfo{author}{\bibfnamefont{X.}~\bibnamefont{{Zhou}}},
  \bibinfo{author}{\bibfnamefont{S.}~\bibnamefont{{Jin}}}, \bibnamefont{and}
  \bibinfo{author}{\bibfnamefont{J.}~\bibnamefont{{Schmiedmayer}}},
  \bibinfo{journal}{New Journal of Physics} \textbf{\bibinfo{volume}{20}},
  \bibinfo{eid}{055005} (\bibinfo{year}{2018}).

\bibitem[{\citenamefont{Bongs et~al.}(1999{\natexlab{b}})\citenamefont{Bongs,
  Burger, Birkl, Sengstock, Ertmer, Rza\ifmmode \mbox{\c{}}\else
  \c{}\fi{}z\ifmmode~\dot{}\else \.{}\fi{}ewski, Sanpera, and
  Lewenstein}}]{Bongs}
\bibinfo{author}{\bibfnamefont{K.}~\bibnamefont{Bongs}},
  \bibinfo{author}{\bibfnamefont{S.}~\bibnamefont{Burger}},
  \bibinfo{author}{\bibfnamefont{G.}~\bibnamefont{Birkl}},
  \bibinfo{author}{\bibfnamefont{K.}~\bibnamefont{Sengstock}},
  \bibinfo{author}{\bibfnamefont{W.}~\bibnamefont{Ertmer}},
  \bibinfo{author}{\bibfnamefont{K.}~\bibnamefont{Rza\ifmmode \mbox{\c{}}\else
  \c{}\fi{}z\ifmmode~\dot{}\else \.{}\fi{}ewski}},
  \bibinfo{author}{\bibfnamefont{A.}~\bibnamefont{Sanpera}}, \bibnamefont{and}
  \bibinfo{author}{\bibfnamefont{M.}~\bibnamefont{Lewenstein}},
  \bibinfo{journal}{Phys. Rev. Lett.} \textbf{\bibinfo{volume}{83}},
  \bibinfo{pages}{3577} (\bibinfo{year}{1999}{\natexlab{b}}),
  \urlprefix\url{https://link.aps.org/doi/10.1103/PhysRevLett.83.3577}.

\bibitem[{\citenamefont{de~Saint-Vincent
  et~al.}(2010)\citenamefont{de~Saint-Vincent, Brantut, Bord{\'{e}}, Aspect,
  Bourdel, and Bouyer}}]{Robert_de_Saint_Vincent_2010}
\bibinfo{author}{\bibfnamefont{M.~R.} \bibnamefont{de~Saint-Vincent}},
  \bibinfo{author}{\bibfnamefont{J.-P.} \bibnamefont{Brantut}},
  \bibinfo{author}{\bibfnamefont{C.~J.} \bibnamefont{Bord{\'{e}}}},
  \bibinfo{author}{\bibfnamefont{A.}~\bibnamefont{Aspect}},
  \bibinfo{author}{\bibfnamefont{T.}~\bibnamefont{Bourdel}}, \bibnamefont{and}
  \bibinfo{author}{\bibfnamefont{P.}~\bibnamefont{Bouyer}},
  \bibinfo{journal}{{EPL} (Europhysics Letters)} \textbf{\bibinfo{volume}{89}},
  \bibinfo{pages}{10002} (\bibinfo{year}{2010}),
  \urlprefix\url{https://doi.org/10.1209/0295-5075/89/10002}.

\bibitem[{\citenamefont{Strecker et~al.}(2002)\citenamefont{Strecker,
  Partridge, Truscott, and Hulet}}]{Strecker}
\bibinfo{author}{\bibfnamefont{K.~E.} \bibnamefont{Strecker}},
  \bibinfo{author}{\bibfnamefont{G.~B.} \bibnamefont{Partridge}},
  \bibinfo{author}{\bibfnamefont{A.~G.} \bibnamefont{Truscott}},
  \bibnamefont{and} \bibinfo{author}{\bibfnamefont{R.~G.} \bibnamefont{Hulet}},
  \bibinfo{journal}{Nature} \textbf{\bibinfo{volume}{417}},
  \bibinfo{pages}{150} (\bibinfo{year}{2002}).

\bibitem[{\citenamefont{Nguyen et~al.}(2014)\citenamefont{Nguyen, Dyke, Luo,
  Malomed, and Hulet}}]{Nguyen}
\bibinfo{author}{\bibfnamefont{J.~H.~V.} \bibnamefont{Nguyen}},
  \bibinfo{author}{\bibfnamefont{P.}~\bibnamefont{Dyke}},
  \bibinfo{author}{\bibfnamefont{D.}~\bibnamefont{Luo}},
  \bibinfo{author}{\bibfnamefont{B.~A.} \bibnamefont{Malomed}},
  \bibnamefont{and} \bibinfo{author}{\bibfnamefont{R.~G.} \bibnamefont{Hulet}},
  \bibinfo{journal}{Nature Physics} \textbf{\bibinfo{volume}{10}},
  \bibinfo{pages}{918} (\bibinfo{year}{2014}).

\bibitem[{\citenamefont{Mewes et~al.}(1997)\citenamefont{Mewes, Andrews, Kurn,
  Durfee, Townsend, and Ketterle}}]{Mewes}
\bibinfo{author}{\bibfnamefont{M.-O.} \bibnamefont{Mewes}},
  \bibinfo{author}{\bibfnamefont{M.~R.} \bibnamefont{Andrews}},
  \bibinfo{author}{\bibfnamefont{D.~M.} \bibnamefont{Kurn}},
  \bibinfo{author}{\bibfnamefont{D.~S.} \bibnamefont{Durfee}},
  \bibinfo{author}{\bibfnamefont{C.~G.} \bibnamefont{Townsend}},
  \bibnamefont{and} \bibinfo{author}{\bibfnamefont{W.}~\bibnamefont{Ketterle}},
  \bibinfo{journal}{Phys. Rev. Lett.} \textbf{\bibinfo{volume}{78}},
  \bibinfo{pages}{582} (\bibinfo{year}{1997}),
  \urlprefix\url{https://link.aps.org/doi/10.1103/PhysRevLett.78.582}.

\bibitem[{\citenamefont{Santos et~al.}(2001)\citenamefont{Santos, Floegel,
  Pfau, and Lewenstein}}]{Santos}
\bibinfo{author}{\bibfnamefont{L.}~\bibnamefont{Santos}},
  \bibinfo{author}{\bibfnamefont{F.}~\bibnamefont{Floegel}},
  \bibinfo{author}{\bibfnamefont{T.}~\bibnamefont{Pfau}}, \bibnamefont{and}
  \bibinfo{author}{\bibfnamefont{M.}~\bibnamefont{Lewenstein}},
  \bibinfo{journal}{Phys. Rev. A} \textbf{\bibinfo{volume}{63}},
  \bibinfo{pages}{063408} (\bibinfo{year}{2001}),
  \urlprefix\url{https://link.aps.org/doi/10.1103/PhysRevA.63.063408}.

\end{thebibliography}


\end{document}